\documentclass[onecolumn,11pt,showpacs,nofootinbib,notitlepage,floatfix,superscriptaddress,aps]{revtex4-1} 
\usepackage{graphicx}
\usepackage{amssymb}
\usepackage{amsmath}
\usepackage{amsthm}
\usepackage{amsfonts}
\usepackage{hyperref}
\usepackage[ruled]{algorithm2e}
\usepackage{diagbox}
\usepackage{mathtools}
 
\usepackage{tikz-cd}
\usepackage{stackengine}

\usepackage{xcolor}
\usepackage{soul}

\newlength{\fighskip} \fighskip=2pt
\newlength{\figvskip} \figvskip=3pt

\newcommand*{\figbox}[2]{{\def\figscale{#1}\def\arraystretch{0.8}\arraycolsep=0pt\begin{array}{c}\vbox{\vskip\figscale\figvskip\hbox{\hskip\figscale\fighskip\includegraphics[scale=\figscale]{#2}}}\end{array}}}

\DeclareMathOperator{\im}{im}
\DeclareMathOperator{\supp}{supp}
\newcommand{\ket}[1]{|#1\rangle}
\newcommand{\bnd}[2]{\partial_{#1,#2}}

\newcommand{\face}[2]{{\Delta_{#1}(#2)}}
\newcommand{\link}[2]{{\mathrm{Lk}_{#1}(#2)}}
\renewcommand{\star}[2]{{\mathrm{St}_{#1}(#2)}}
\newcommand{\col}[1]{{\mathrm{col}(#1)}}
\newcommand{\com}[1]{K[#1]}
\newcommand{\sini}{\mathcal{S}^Z_{\textrm{ini}}}
\newcommand{\sfin}{\mathcal{S}^X_{\textrm{fin}}}
\newcommand{\hini}{\mathcal{H}(\mathcal{S}_{\textrm{ini}}^Z)}
\newcommand{\hfin}{\mathcal{H}(\mathcal{S}_{\textrm{fin}}^X)}

\newcommand{\ugmap}{\quad\xmapsto{\widetilde\Gamma}\quad} 

\begin{document}

\title{Ungauging quantum error-correcting codes}
\author{Aleksander Kubica}
\affiliation{Perimeter Institute for Theoretical Physics, Waterloo, Ontario N2L 2Y5, Canada}
\affiliation{Institute for Quantum Computing, University of Waterloo, Waterloo, ON N2L 3G1, Canada}
\author{Beni Yoshida}
\affiliation{Perimeter Institute for Theoretical Physics, Waterloo, Ontario N2L 2Y5, Canada}
\date{\today}

\begin{abstract}

We develop the procedures of gauging and ungauging, reveal their operational meaning and propose their generalization in a systematic manner within the framework of quantum error-correcting codes.
We demonstrate with an example of the subsystem Bacon-Shor code that the ungauging procedure can result in models with unusual symmetry operators constrained to live on lower-dimensional structures.
We apply our formalism to the three-dimensional gauge color code (GCC) and show that its codeword space is equivalent to the Hilbert space of six copies of $\mathbb{Z}_2$ lattice gauge theory with $1$-form symmetries.
We find that three different stabilizer Hamiltonians associated with the GCC correspond to distinct thermal symmetry-protected topological (SPT) phases in the presence of the stabilizer symmetries of the GCC.
One of the considered Hamiltonians describes the Raussendorf-Bravyi-Harrington model, which is universal for measurement-based quantum computation at non-zero temperature.
We also propose a general procedure of creating $D$-dimensional SPT Hamiltonians from $(D+1)$-dimensional CSS stabilizer Hamiltonians by exploiting a relation between gapped domain walls and transversal logical gates.
As a result, we find an explicit two-dimensional realization of a non-trivial fracton SPT phase protected by fractal-like symmetries.

\end{abstract}

\maketitle

\section{Introduction}

Duality maps have proven to be a powerful tool in theoretical physics, which allow us to relate two seemingly different systems by transforming observables in one of them into the observables in the other while preserving their algebraic structure.
Gauging is a particular form of a duality map which couples a system with global symmetries associated with the symmetry group $G$ to gauge fields with the same symmetry~\cite{Kogut:1979aa, Wegner:1971aa, Fradkin:1979aa}. 
Gauging transforms the structure of entanglement in a non-trivial manner by mapping trivial short-range entangled states into the gauged states, which usually possess long-range entanglement such as topological order.
The duality relation via gauging has provided a broad framework to address questions related to the classification of topological phases of matter and has led to the discovery of exotic phases~\cite{Dijkgraaf90, Kitaev03, Wen_Text, Chen13, Levin12}.
Recently, the procedures of gauging and ungauging have been used to address certain fundamental questions in quantum information theory such as the classification of transversal logical gates and resources required for universal quantum computation~\cite{Else12, Beni15c, Roberts:2017ab, Nautrup:2015aa, Miller:2016aa}.

Despite these successes, the definition and operational interpretation of gauging still remain elusive, which can be partly attributed to the fact 
that not all physical operators are allowed due to the symmetry constraints.
Ambiguities in gauging have led to serious confusions and debates over factorization of Hilbert spaces in the presence of symmetries, see for instance~\cite{Casini:2014aa, Soni:2016aa, Van-Acoleyen:2016aa} and references therein. 
Moreover, while the procedure of ungauging the symmetries of a theory can be defined in an unambiguous manner, the reverse process may not be unique~\cite{Jenkins:2013aa, Haegeman15}.
The goal of this paper is thus to revisit the procedure of gauging, reveal its operational meaning and propose its generalization in a systematic manner through the lens of theory of quantum error correction.

Recently several gauging procedures have been discussed in the context of classification of topological phases of matter~\cite{Kapustin13, Gaiotto15, Beni15b, Vijay:2016aa, Williamson:2016aa}.
While these approaches may differ in some details, the underlying structure bears certain similarities. 
First of all, gauging can be defined as an isomorphism, i.e.., a one-to-one map, between symmetric subspaces $\mathcal{H}(\mathcal{S}_{\textrm{ini}})$ and $\mathcal{H}(\mathcal{S}_{\textrm{fin}})$ of two Hilbert spaces $\mathcal{H}_{\textrm{ini}}$ and $\mathcal{H}_{\textrm{fin}}$.
The subspace $\mathcal{H}(\mathcal{S}_{\textrm{ini}})$ is defined via the global symmetry group $\mathcal{S}_{\textrm{ini}}$, whereas the other subspace $\mathcal{H}(\mathcal{S}_{\textrm{fin}})$ is determined by the local symmetry group $\mathcal{S}_{\textrm{fin}}$.
The symmetry operators from $\mathcal{S}_{\textrm{ini}}$ are mapped to the identity operator via the gauging map whereas new gauge symmetry operators (forming a group $\mathcal{S}_{\textrm{fin}}$) emerge in order to consistently define the gauging map.
Second, most of the aforementioned (generalized) gauging procedures transform a trivial ground state of the Ising paramagnet into states which are the codewords of some special class of quantum error-correcting codes, i.e., Calderbank-Shor-Steane (CSS) stabilizer codes.
This observation motivates us to further explore and develop a framework for gauging and ungauging applicable to CSS codes.

To describe the procedures of gauging and ungauging, we use a notion of a chain complex.
A chain complex is a sequence of vector spaces and linear maps between them, called boundary operators, such that the composition of two consecutive maps is the zero map.
Importantly, CSS stabilizer codes can also be described using the formalism of chain complexes~\cite{Bravyi_Hastings13, Bombin_review}.
This leads to the intimate connection between the CSS codes and the systematic procedure of gauging (and ungauging) stabilizer symmetries.

In this paper, we will be primarily interested in explorations of the procedures of gauging and ungauging of Pauli CSS symmetries for discrete systems of qubits (or qudits in general). 
We will apply the procedures of gauging and ungauging to various quantum error-correcting codes and investigate corresponding quantum phases of matter.
In what follows, we briefly describe our results, which include: $(i)$ a systematic framework for gauging and ungauging accociated with the CSS stabilizer and subsystem codes, $(ii)$ a study of the three-dimensional gauge color code and its thermal stability in the presence of stabilizer symmetries, and $(iii)$ a definition of a new phase, the fractal symmetry-protected topological phase, arising in the presence of fractal-like symmetries.

\subsection*{Ungauging subsystem codes}

The procedures of gauging and ungauging prove to be particularly powerful in characterizing the codeword space of subsystem codes.
Subsystem codes are a natural generalization of stabilizer codes, since we do not require that their code generators commute~\cite{Poulin:2005aa, Kribs06, Bacon06, Bravyi10}. 
Intuitively, one may interpret a subsystem code as a stabilizer code, where quantum information is encoded only into a subset of the logical qubits. 
The codeword space of a subsystem code can be extensive in the system size which provides certain advantages over stabilizer codes.
For instance, by fixing the state of the unused logical qubits, one may be able to fault-tolerantly switch between different stabilizer codes~\cite{Paetznick13}.
Also, the syndrome extraction may become simpler due to reduced weight of Pauli operators required to measure. 
From condensed matter viewpoint, subsystem codes are likely to host a wide variety of interesting quantum phases and some exactly solvable Hamiltonians~\cite{Xu04, Kitaev06b}.
In fact, a subsystem code does not unambiguously define a subsystem Hamiltonian.
Rather, it gives rise to multiple gapped stabilizer Hamiltonians as well as frustrated gapless Hamiltonians, possibly at a quantum phase transition. 

Our current understanding of subsystem codes remains limited due to the very fact that the codeword space may be extensive and possibly contains codeword spaces of multiple stabilizer codes.
At the moment, a unified framework to study a family of stabilizer Hamiltonians arising from subsystem codes as well as the structure of their codeword space is missing.
In Sec.~\ref{sec_ungauging}, we develop the ungauging procedure applicable to the class of CSS subsystem codes, whose code generators consist only of $X$-type and $Z$-type Pauli operators.
By ungauging $Z$-type symmetries while preserving $X$-type symmetries, the codeword space of a subsystem code is mapped to a subspace of the Hilbert space determined by only $X$-type symmetries, and thus is easier to analyze. 

We shall demonstrate the usefulness of this approach by ungauging the 2D subsystem Bacon-Shor code~\cite{Bacon06}. 
We remark that in a typical scenario the emergent global symmetries act on the entire system as a tensor product of local operators as in the case of the $\mathbb{Z}_2$ paramagnet.
Moreover, ordinary higher-form symmetries act on subvolumes of the system and their geometric shapes can be continuously deformed. 
By ungauging the 2D Bacon-Shor code we find unusual symmetry operators which, unlike the conventional symmetries, are constrained to lower-dimensional submanifolds and are not deformable, similar to the ones discussed in~\cite{Vijay:2016aa, Williamson:2016aa}.
This finding of rigid symmetry operators naturally suggests possibility of novel quantum phases of matter.
Indeed, such symmetries have recently been identified and dubbed subsystem symmetries in the independent work~\cite{You18}.
Also, the appearance of such exotic symmetries naturally leads us to the question of an appropriate gauging procedure for lower-dimensional rigid symmetries.
The answer can be unambiguously obtained by simply reversing the procedure of ungauging, which corresponds to eliminating a certain subgroup of the total symmetry group.
In this sense, the gauging procedure to obtain the codeword Hilbert space of a CSS subsystem code can be interpreted as a partial gauging of a standard Hilbert space with $X$-type symmetries.

\subsection*{Ungauging the 3D gauge color code}

We use the ungauging procedure to study the 3D gauge color code (GCC), which is an example of the 3D subsystem topological code~\cite{Bombin15}.
The GCC is a particularly promising model for fault-tolerant quantum computation \cite{Kubicathesis, Brown2015, Kubica2017, Sam_Stephen}.
Namely, one can achieve universality with the color code by switching between two versions of the code, which correspondingly admit transversal logical Clifford operators and a logical non-Clifford $T$ gate~\cite{Paetznick13, Kubica15, Bombin:2016aa}.
{This in turn may lead to a computational scheme with reduced resource overhead compared to standard methods.}

Despite these promising features, the properties of the GCC are not fully understood.
For instance, since single-shot error-correction~\cite{Bombin:2015aa} is a property typically exhibited by self-correcting quantum memories (such as the 4D toric code), it has been suspected that the GCC also possesses some sort of thermal topological order and stability.
However, none of stabilizer codes, including the 3D stabilizer color code, whose codeword spaces can be reached from the code subspace of the GCC, are topologically ordered at nonzero temperature. 
At the same time,  a symmetry-protected topological (SPT) phase with a surface topological order can appear in the 3D GCC with boundaries. 
Understanding the structure of quantum phases arising from the 3D GCC is an interesting step to unveil a connection between fault-tolerant quantum computation and quantum many-body physics.

In Sec.~\ref{sec_gcc}, by ungauging the $Z$-type stabilizer symmetries of the GCC, we show that the codeword space of the GCC is equivalent to the Hilbert space of lattice gauge theory with six copies of $\mathbb{Z}_{2}$ $1$-form symmetries. 
Moreover, we find that different stabilizer Hamiltonians associated with the GCC correspond to different fixed-point Hamiltonians in the presence of the aforementioned symmetries.
In particular, one of the ungauged Hamiltonians is equivalent to the Raussendorf-Bravyi-Harrington (RBH) model which is universal for measurement-based quantum computation at nonzero temperature~\cite{Raussendorf:2005aa}.
We remark that while the RBH model is thermodynamically trivial, it possesses thermal order in the presence of $1$-form symmetries~\cite{Beni15b, Roberts:2017ab}, which intuitively explains its high thresholds for fault-tolerant quantum computation~\cite{, Raussendorf:2006aa}.
Our results present a possible answer to the question of thermal stability of the GCC.
Namely, the stabilizer Hamiltonians identified within the codeword space of the GCC are thermally stable in the presence of the stabilizer symmetries.
Moreover, the ground states and the thermal Gibbs ensembles of these stabilizer Hamiltonians belong to different phases of matter and their phase separation survives even at non-zero temperature as long as the stabilizer symmetries are enforced.

\subsection*{Fracton SPT phases}

Fracton codes \cite{Newman99,  Castelnovo12, Haah11, Beni11b,Beni13, Vijay:2015aa} are a particularly intriguing class of 3D geometrically-local stabilizer codes, whose logical operators form fractals, and thus are beyond the description of a standard topological field theory.
Our ungauging procedure based on the chain complex can be readily applied to ungauge fractal-like symmetries of any CSS fracton code.
In particular, in Sec.~\ref{sec_fracSPT} we focus on ungauging a simple 3D model with a gapped domain wall, which is constructed from the 3D fractal code~\cite{Beni11b} with logical operators in the shape resembling the Sierpi\'nski triangle.
As a result, we obtain a non-trivial example of a new phase of matter, which we call a fractal symmetry-protected topological (frac-SPT) phase.
We remark that we use the abbreviation ``frac-SPT'' since fSPT is often used for Floquet SPT phases.
The non-triviality of the aforementioned model of fracton SPT can be shown by utilizing a connection between gapped domain walls and transversal logical operators~\cite{Beni15, Beni15c}.
In fact, our approach is quite general --- given a $(D+1)$-dimensional CSS stabilizer code with geometrically-local generators, one can construct a $D$-dimensional SPT Hamiltonian which is realized as a $D$-dimensional gapped domain wall of the $(D+1)$-dimensional model. 
Our approach not only provides a systematic way of generating SPT Hamiltonians but also suggests their physical realization as gapped domain walls.

\subsection*{Miscellaneous results}

In addition to the aforementioned main contributions of the paper, there are two other results which we relegate to the appendix. 

\begin{itemize}

\item{\textbf{Equivalence of the color and toric codes:}}
The $D$-dimensional color code has been previously shown to be equivalent to multiple copies of the $D$-dimensional toric code up to a geometrically-local unitary circuit of constant depth~\cite{Kubica15b}.
We find that this unitary mapping can be viewed as ungauging of certain symmetries of the color code.
This is discussed in Appendix \ref{sec_equiv}.

\item{\textbf{3D Majorana gauge color code:}}
Since the 3D gauge color code is self dual, i.e., its code generators are invariant under the exchange of $X$ and $Z$ operators, one can construct its variant consisting of Majorana fermion operators.
We identify two particularly interesting 3D Majorana Hamiltonians associated with the 3D Majorana gauge color code.
The first model supports the 2D Majorana color code on the surface while having a trivial bulk.
The second model possesses 3D topological order and supports fermionic point-like excitations.
This is discussed in Appendix \ref{sec:majorana}.
\end{itemize}

\section{Ungauging subsystem codes}
\label{sec_ungauging}

In this section we focus on the procedures of gauging and ungauging the Pauli CSS symmetries.
{The ungauging procedure transforms between symmetric subspaces $\hini$ and $\hfin$ of two Hilbert spaces $\mathcal{H}_{\textrm{ini}}$ and $\mathcal{H}_{\textrm{fin}}$, where $\hini$ and $\hfin$ are determined by the initial $Z$-type $\sini$ and the emergent $X$-type $\sfin$ symmetry groups.}
In Sec.~\ref{sec_definition} we define the ungauging map $\widetilde\Gamma$ based on some chain complex.
Then, we discuss the operational meaning of gauging and ungauging.
We illustrate our ungauging procedure with two examples, in which we ungauge the (stabilizer) toric code and the (subsystem) Bacon-Shor code. 
Finally, we discuss an issue of partial gauging of $X$-type symmetries, which arises in the context of CSS subsystem codes.

\subsection{Basics of CSS stabilizer codes}

Let us consider a system composed of qubits, which are labeled by the elements from the set $\mathcal{B}_Q$.
We write $X_i$ or $Z_i$ to represent a Pauli $X$ or $Z$ operator supported on the qubit $i \in \mathcal{B}_Q$.
We say that an operator is of $X$-type or $Z$-type if it is a product of either Pauli $X$ or Pauli $Z$, respectively.
We denote by $X(\mathcal{I}) = \prod_{i \in \mathcal{I}} X_i$ an $X$-type operator supported on the subset of qubits $\mathcal{I} \subset \mathcal{B}_Q$; similarly $Z(\mathcal{I})$.
We can express the commutation relation between $X(\mathcal{I}_1)$ and $Z(\mathcal{I}_2)$ as follows
\begin{equation}
\com{X(\mathcal{I}_1),Z(\mathcal{I}_2)} = (-1)^{|\mathcal{I}_1 \cap \mathcal{I}_2|} I,
\end{equation}
where $\com{U_1,U_2} = U_1 U_2 U^\dag _1 U^\dag _2$ is the group commutator of two operators $U_1$ and $U_2$.
In other words, $X(\mathcal{I}_1)$ and $Z(\mathcal{I}_2)$ anticommute iff they overlap on an odd number of qubits, i.e., $|\mathcal{I}_1 \cap \mathcal{I}_2| \equiv 1 \mod 2$.

A stabilizer code \cite{Gottesman96} is defined by a stabilizer group $\mathcal{S}_{\textrm{stab}}$, which is an Abelian subgroup of the Pauli group $\mathcal{P}$ and does not contain $-I$.
The code space is the $(+1)$-eigenspace of all elements of the stabilizer group.
Importantly, the code space can be viewed as the energy ground space of the stabilizer Hamiltonian
which is the sum of stabilizers generating the stabilizer group $\mathcal{S}_{\textrm{stab}}$.
The unitary operators which preserve the codespace are called logical operators.

CSS stabilizer codes \cite{Calderbank96}, which include the well-known toric code and the color code~\cite{Bombin06, Bombin07}, are a special class of stabilizer codes.
The stabilizer group of a CSS code is generated by Pauli operators, which can be chosen to be either $X$-type or $Z$-type but not mixed.
For any CSS stabilizer code we can introduce a CSS chain complex
\begin{equation}
\begin{tikzcd}
\raisebox{-8pt}{\stackanchor[8pt]{$C_Z$}{$Z$-stabilizers}} \arrow[r,"\partial_Z"] &
\raisebox{-8pt}{\stackanchor[8pt]{$C_Q$}{qubits}} \arrow[r,"\partial_X"] &
\raisebox{-8pt}{\stackanchor[8pt]{$C_X$}{$X$-stabilizers}}
\end{tikzcd}
\label{eq_chain_CSS}
\end{equation}
where $C_Z$, $C_Q$ and $C_X$ are finitel-dimensional $\mathbb{F}_2$-vector spaces with bases $\mathcal{B}_Z = Z\textrm{-stabilizer}$ generators, $\mathcal{B}_Q = \textrm{qubits}$ and $\mathcal{B}_X = X\textrm{-stabilizer}$ generators.
The linear maps $\partial_Z$ and $\partial_X$, called boundary operators, 
can be viewed as binary matrices of size $\dim C_Q \times \dim C_Z$ and $\dim C_X \times \dim C_Q$ corresponding to the parity-check matrices $H^T_Z$ and $H_X$ of the underlying CSS code.
The boundary operators are defined as follows: the $i$th column of $\partial_Z$ corresponds to the support of the $Z$-type stabilizer labeled by $i\in \mathcal{B}_Z$, and the $j$th row of $\partial_X$ corresponds to the support of the $X$-type stabilizer labeled by $j\in \mathcal{B}_X$.
By definition, the composition of boundary operators is the zero map.
This condition can be phrased as a product of two matrices $\partial_X$ and $\partial_Z$ being zero, $\partial_X \cdot \partial_Z = 0$.

We can consider other linear maps associated with the CSS chain complex in Eq.~(\ref{eq_chain_CSS}).
Namely, we define the coboundary operators
\footnote{Technically, the coboundary operators are defined between the dual spaces $C_X^*$, $C_Q^*$ and $C_Z^*$.
However, since $C_X^*$, $C_Q^*$ and $C_Z^*$ are isomorphic to $C_X$, $C_Q$ and $C_Z$, thus we will treat the coboundary operators as maps between $C_X$, $C_Q$ and $C_Z$.}
$\partial_X^T: C_X \rightarrow C_Q$ and $\partial_Z^T: C_Q \rightarrow C_Z$
via binary matrices $(\partial_X)^T$ and $(\partial_Z)^T$, where $M^T$ denotes the transpose of the matrix $M$.
The composition of coboundary operators is the zero operator, which can be stated as $(\partial_Z)^T\cdot (\partial_X)^T = 0$.

We remark that the support of any $Z$-type $c_Z \in C_Z$ and $X$-type $c_X \in C_X$ stabilizers is $\partial_Z c_C$ and $\partial^T_X c_X$.
Moreover, Pauli $X$-type and $Z$-type errors are detected by appropriate stabilizer measurements.
Namely, a Pauli $Z$-type error $Z(c_Q)$ leads to an $X$-type syndrome pattern $\partial_{X} c_{Q}$, whereas Pauli $X$-type error $X(c_Q)$ returns a $Z$-type syndrome $\partial_{Z}^{T} c_{Q}$, where $c_Q \in C_Q$ represents any subset of qubits.
Lastly, non-trivial Pauli $Z$ and $X$ logical operators can be found as elements of respectively $\ker \partial_X \setminus \im \partial_Z$ and $\ker \partial^T_Z \setminus \im \partial^T_X$ since logical operators must have trivial error syndromes.

\begin{figure}[h!]
\centering
\includegraphics[width=.5\columnwidth]{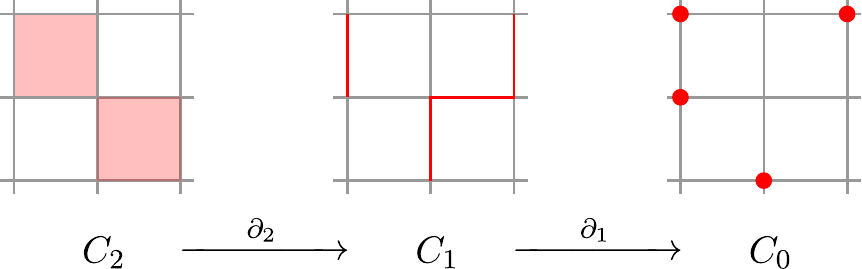}
\caption{
A chain complex associated with a 2D square lattice $\mathcal{L}$.
The $\mathbb{F}_2$-vector spaces $C_2$, $C_1$, and $C_0$ are associated with faces, edges and vertices of $\mathcal{L}$.
The boundary operator $\partial_2 : C_2 \rightarrow C_1$ is a linear map, which for each face returns the edges around that face.
Similarly, $\partial_1: C_1 \rightarrow C_0$ is a linear map, which returns the vertices belonging to the given edge.
We remark that binary vectors $c_i \in C_i$ (shaded in red) can be thought of as subsets of $i$-dimensional elements of the lattice $\mathcal{L}$.
}
\label{fig_chain_complex} 
\end{figure}

To illustrate the notion of a CSS chain complex, let us consider the toric code defined on a two-dimensional lattice $\mathcal{L}$.
We denote the set of $i$-cells of $\mathcal{L}$ by $\face i {\mathcal{L}}$.
In particular, $\face 0 {\mathcal{L}}$, $\face 1 {\mathcal{L}}$ and $\face 2 {\mathcal{L}}$ correspond to vertices, edges and faces of $\mathcal{L}$.
The CSS chain complex of the toric code coincides with the standard chain complex 
$C_2 \xrightarrow{\partial_2} C_1 \xrightarrow{\partial_1} C_0$
associated with the lattice $\mathcal{L}$.
Namely, the bases for $\mathbb{F}_2$-vector spaces $C_2$, $C_1$ and $C_0$ are the sets of faces, edges and vertices, $\mathcal{B}_i = \face i {\mathcal{L}}$, and the boundary operator $\partial_i$ returns all the $(i-1)$-cells belonging to the given $i$-cell; see Fig.~\ref{fig_chain_complex} for a schematic illustration.

We remark that the toric code can be defined in any dimensions by considering a $D$-dimensional lattice $\mathcal{L}$, where $D \geq 2$.
The toric code of type $k \in \{1,\ldots, D-1 \}$ is defined by placing qubits on the $k$-dimensional cells $\face k {\mathcal{L}}$  and associating $X$- and $Z$-type stabilizer generators with $(k-1)$- and $(k+1)$-cells.
The CSS chain complex coincides with the part 
$C_{k+1} \xrightarrow{\partial_{k+1}} C_k \xrightarrow{\partial_k} C_{k-1}$
of the standard chain complex for $\mathcal{L}$.
Similarly, we can introduce a CSS chain complex associated with the $d$-dimensional stabilizer color code.
However, due to technical complications, such as a need for the generalized boundary operator $\bnd k l$, we defer further discussion of the color code to Sec.~\ref{sec_GCC_def} and Appendix~\ref{sec_equiv}.

\subsection{Ungauging based on a chain complex}
\label{sec_definition}

\begin{figure}[h!]
\centering
\includegraphics{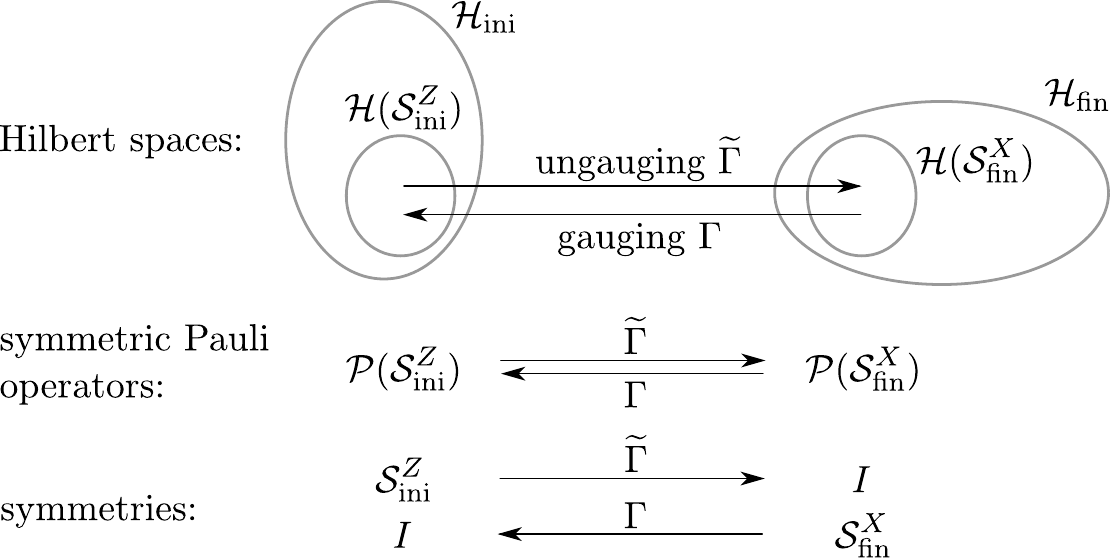}
\caption{
The ungauging map $\widetilde\Gamma$ and the gauging map $\Gamma$ are isomorphisms between symmetric subspaces $\hini$ and $\hfin$ of two Hilbert spaces $\mathcal{H}_{\textrm{ini}}$ and $\mathcal{H}_{\textrm{fin}}$.
The subspaces $\hini$ and $\hfin$ are defined by the initial $Z$-type $\sini$ and the emergent $X$-type $\sfin$ symmetry groups.
The ungauging map $\widetilde\Gamma$ transforms operators from the symmetric Pauli subgroup $\mathcal{P}(\sini)$ into operators in $\mathcal{P}(\sini)$; the gauging map $\Gamma$ can be viewed as an inverse of $\widetilde\Gamma$.
Ungauging eliminates the initial $Z$-type symmetry group $\sini$, whereas gauging eliminates the emergent $X$-type symmetry group $\sfin$.
}
\label{fig_ungauging_Hilbert} 
\end{figure}

The ungauging procedure allows us to map between subspaces $\hini$ and $\hfin$ of two Hilbert spaces $\mathcal{H}_{\textrm{ini}}$ and $\mathcal{H}_{\textrm{fin}}$.
The Hilbert spaces $\mathcal{H}_{\textrm{ini}}$ and $\mathcal{H}_{\textrm{fin}}$ are identified with the initial $Q_{\textrm{ini}}$ and the final $Q_{\textrm{fin}}$ sets of qubits, respectively.
By convention, the subspace $\hini$ is defined via the initial $Z$-type symmetry group $\sini$, where $\sini$ is a subgroup of the Pauli group $\mathcal{P}_{\textrm{ini}}$.
Namely, $\hini$ is the subspace of $\mathcal{H}_{\textrm{ini}}$ spanned by the $(+1)$-eigenstates states of the symmetry operators from $\sini$, namely
\begin{equation}
\hini = \{ \ket{\psi} \in \mathcal{H}_{\textrm{ini}} | \forall S^Z \in \sini: S^Z \ket{\psi} = + \ket{\psi} \}.
\end{equation}
We label the $Z$-type generators $S^Z_j \in \mathcal{P}_{\textrm{ini}}$ of the symmetry group $\sini$ by the elements $j$ of the set $\mathcal{B}_Z$, and thus have $\sini = \langle S^Z_j | j \in \mathcal{B}_Z \rangle$.
Similarly, the subspace $\hfin$ of the Hilbert space $\mathcal{H}_{\textrm{fin}}$ is defined by the emergent $X$-type symmetry group $\sfin$, which is a subgroup of the Pauli group $\mathcal{P}_{\textrm{fin}}$.
We schematically depict the ungauging procedure in Fig.~\ref{fig_ungauging_Hilbert}.

The ungauging map $\widetilde\Gamma$ is an isomorphism between $\hini$ and $\hfin$, and thus is defined for all states and operators on $\hini$.
Since $\widetilde\Gamma$ is a linear map, it can be specified by its action on the elements of some basis of operators, for instance 
the symmetric subgroup $\mathcal{P}(\sini)$ of the Pauli group $\mathcal{P}_{\textrm{ini}}$.
The symmetric Pauli subgroup $\mathcal{P}(\sini)$ is defined as the group of Pauli operators, which commute with the symmetry operators from $\sini$, namely
\begin{equation}
\mathcal{P}(\sini) = \{ P \in \mathcal{P}_{\textrm{ini}} | \forall S^Z \in \sini: \com{P, S^Z} = I \}.
\end{equation}
Importantly, the ungauging map has to satisfy two conditions
\begin{itemize}
\item[$(i)$] any $Z$-type symmetry operator $S^Z \in\sini$ is mapped to the identity operator, i.e.,
$S^Z \xmapsto{\widetilde\Gamma} I$,
\item[$(ii)$] the inner product between any two states $\ket {\psi_1}, \ket {\psi_2} \in \hini$ is preserved by $\widetilde\Gamma$.
\end{itemize}
We remark that the condition $(ii)$ can be verified by checking whether the commutation relation between any two elements of the symmetric Pauli subgroup $\mathcal{P}(\sini)$ is preserved by $\widetilde\Gamma$.

The first step of the ungauging procedure is to identify the generators of the symmetric Pauli subgroup $\mathcal{P}(\sini)$.
We observe that $\mathcal{P}(\sini)$ is generated by single-qubit $Z$, as well as by some (possibly multi-qubit) $X$-type operators $P^X_k \in \mathcal{P}_{\textrm{ini}}$ labeled by $k\in\mathcal{B}_X$.
Namely,
\begin{equation}
\mathcal{P}(\sini) = \langle Z_i, P^X_k | i\in\mathcal{B}_Q, k \in \mathcal{B}_X \rangle,
\end{equation}
where we relabel $Q_{\textrm{ini}}$ as $\mathcal{B}_Q$ for notational convenience.
The $X$-type generators $P^X_k$ of the symmetric Pauli group $\mathcal{P}(\sini)$ do not have to be independent, i.e., it might be possible to obtain the identity operator by multiplying some generators $P^X_k$.
We define a relation $R^{(l)}$ to be a binary row vector of length $|\mathcal{B}_X|$ encoding which generators $P^X_k$ are not independent, namely
\begin{equation}
\prod_{\substack{k\in \mathcal{B}_X\\ [R^{(l)}]_k =1 }} P^X_k = I,
\end{equation}
where $[R^{(l)}]_k$ denotes the $k$th entry $R^{(l)}$.
Note that by adding two relations modulo 2 we get a new relation, and thus we can think of the set of all relations as a group.
We label all the relations $\{ R^{(l)} \}$ by the elements $l\in\mathcal{B}_R$.

The next step of the ungauging procedure is to introduce the ungauging chain complex
\begin{equation}
\begin{tikzcd}
\raisebox{-8pt}{\stackanchor[8pt]{$C_Z$}{$Z$-symmetries}} \arrow[r,"\partial_Z"] &
\raisebox{-8pt}{\stackanchor[8pt]{$C_Q$}{initial qubits}} \arrow[r,"\partial_X"] &
\raisebox{-8pt}{\stackanchor[8pt]{$C_X$}{$X$-operators}} \arrow[r,"\partial_R"] &
\raisebox{-8pt}{\stackanchor[8pt]{$C_R$}{$X$-relations}}
\end{tikzcd}
\label{eq_chain_ungauge}
\end{equation}
The chain complex in Eq.~(\ref{eq_chain_ungauge}) is a sequence of $\mathbb{F}_2$-vector spaces $C_Z$, $C_Q$, $C_X$ and $C_R$ with bases 
$\mathcal{B}_Z$, $\mathcal{B}_Q$, $\mathcal{B}_X$ and $\mathcal{B}_R$, respectively.
The boundary operators  are specified by binary matrices $\partial_Z$, $\partial_X$ and $\partial_R$.
Importantly, the composition of two consecutive boundary operators is the zero map, which can be expressed as the matrix relations $\partial_X \cdot \partial_Z = 0$ and $\partial_R \cdot \partial_X = 0$.
We introduce the coboundary operators via the corresponding transposed binary matrices $\partial^T_Z$, $\partial^T_X$ and $\partial^T_R$.
The $j$th column of the matrix $\partial_Z$ represents the support of the generator $S^Z_j$ of the $Z$-type initial symmetry group $\sini$, where $j\in\mathcal{B}_Z$. 
In other words, the matrix element $[\partial_Z]_{i,j} = 1$ iff the qubit $i\in \mathcal{B}_Q$ is in the support of $S^Z_j$, i.e., $i\in \supp S^Z_j$.  
The $k$th row of the matrix $\partial_X$, where $k\in\mathcal{B}_X$, represents the support of the $X$-type generator $P^X_k$ of the symmetric Pauli group $\mathcal{P}(\sini)$, i.e., $[\partial_X]_{k,i} = 1$ iff $i\in \supp P^X_k$.
We remark that any $Z$- and $X$-type symmetric operators from $\mathcal{P}(\sini)$ are of the form $Z(c_Q)$ and $X(\partial^T_X \cdot c_X)$ for some $c_Q \in C_Q$ and $c_X \in C_X$.
Lastly, the $l$th row of the matrix $\partial_R$ is the relation $R^{(l)}$, where $l\in\mathcal{B}_R$.

The ungauging chain complex in Eq.~(\ref{eq_chain_ungauge}) can be unambiguously constructed from the initial Hilbert space $\mathcal{H}_{\textrm{ini}}$ and the initial $Z$-type symmetry group $\sini$ by specifying the $X$-type generators of the symmetric Pauli group $\mathcal{P}(\sini)$ and the relations between them.
Then, the ungauging chain complex determines the final Hilbert space $\mathcal{H}_{\textrm{fin}}$ and the emergent $X$-type symmetries of the subspace $\hfin$ as follows.
For every $X$-type generator $P^X_k \in \mathcal{P}_{\textrm{ini}}$ of the symmetric Pauli subgroup $\mathcal{P}(\sini)$ we introduce one qubit, and thus we identify the final set of qubits $Q_{\textrm{fin}}$ with $\mathcal{B}_X$.
The elements of the emergent $X$-type symmetry group $\sfin$ are chosen to be $X$-type operators of the form $X(M_R \cdot c_R) \in \mathcal{P}_{\textrm{fin}}$ for any $c_R \in C_R$.
One can check that the resulting subspace $\hfin$ is of the same dimension as $\hini$, i.e., $\dim \hfin = \dim \hini$.
Thus, we can find an isomorphism between $\hini$ and $\hfin$, which is the ungauging map $\widetilde\Gamma$.

The ungauging map $\widetilde\Gamma$ between subspaces $\hini$ and $\hfin$ of two Hilbert spaces $\mathcal{H}_{ini}$ and $\mathcal{H}_{fin}$ is determined by its action on the $Z$- and $X$-type operators $Z(c_Q)$ and $X(M_X \cdot c_X)$ from the symmetric Pauli group $\mathcal{P}(\sini)$.
This can be read off from the ungauging chain complex, namely
\begin{eqnarray}
\label{eq_gaugeZ}
\forall c_Q\in C_Q: Z(c_Q) &\quad\xmapsto{\widetilde\Gamma}\quad& Z(\partial_X \cdot c_Q),\\
\label{eq_gaugeX}
\forall c_X\in C_X: X(\partial^T_X \cdot c_X) &\quad\xmapsto{\widetilde\Gamma}\quad& X(c_X).
\end{eqnarray}
We can easily verify that symmetric Pauli operators from $\mathcal{P}(\sini)$ are mapped by $\widetilde\Gamma$ to symmetric Pauli operators from $\mathcal{P}(\sfin)$.
Moreover, we note that $\widetilde\Gamma$ defined by Eqs.~(\ref{eq_gaugeZ})~and~(\ref{eq_gaugeX}) satisfies two conditions $(i)$ and $(ii)$ for the ungauging map.
First, for any $c_Z \in C_Z$ the symmetry operator $Z(\partial_Z \cdot c_Z)$ from the $Z$-type initial symmetry group $\sini$ is mapped by $\widetilde\Gamma$ to the identity operator
\begin{equation}
Z(\partial_Z \cdot c_Z) \xmapsto{\widetilde\Gamma}
Z(\partial_X \cdot (\partial_Z \cdot c_Z)) = Z((\partial_X \cdot \partial_Z) \cdot c_Z)) = I,
\end{equation}
where we use linearity of $\widetilde\Gamma$, Eq.~(\ref{eq_gaugeZ}) and $\partial_X \cdot \partial_Z = 0$.
Second, the commutation relation between any two symmetric Pauli operators is preserved by $\widetilde\Gamma$.
For instance, consider any two operators $Z(c_Q)$ and $X(\partial^T_X\cdot c_X)$ from $\mathcal{P}(\mathcal{S}_Z)$.
They satisfy the commutation relation $[Z(c_Q), X(\partial^T \cdot c_X)] = (-1)^{c_Q^T \cdot (\partial^T_X \cdot c_X)} I$.
After ungauging, we have $[Z(\partial_X \cdot c_Q), X(c_X)] = (-1)^{(\partial^T_X \cdot c_Q )^T\cdot c_X} I$, and thus the commutation relation is the same.
Finally, we conclude that specifying the ungauging chain complex in~Eq.~(\ref{eq_chain_ungauge}) determines the ungauging procedure and provides an explicit recipe for the ungauging map $\widetilde\Gamma$.

\begin{figure}[h!]
\centering
\includegraphics{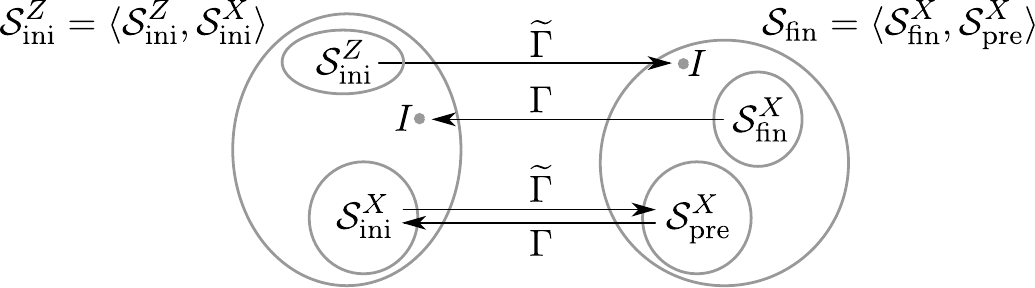}
\caption{
The symmetry groups $\mathcal{S}_{\textrm{ini}}$ and $\mathcal{S}_{\textrm{fin}}$ of the initial and final system.
The ungauging $\widetilde\Gamma$ and $\Gamma$ maps transform correspondingly the initial $Z$-type symmetries $\sini$ and the emergent $X$-type symmetries $\sfin$ into the identity operator.
If there are some additional $X$-type symmetries $\mathcal{S}^X_{\textrm{ini}}$ in the initial system, then they will be preserved and mapped by $\widetilde\Gamma$ into the preserved $X$-type symmetries $\mathcal{S}^X_{\textrm{pre}}$ in the final system.
Similarly, if we gauge the symmetry group $\sfin$ of the final system, then $\Gamma$ would map $\mathcal{S}^X_{\textrm{pre}}$ into $\mathcal{S}^X_{\textrm{ini}}$ in the initial system.
}
\label{fig_ungauging_symmetries} 
\end{figure}

We remark that the ungauging procedure we just described maps between the subspace $\hini$ determined by the $Z$-type initial symmetry group $\sini$ and the subspace $\hfin$ determined by the $X$-type emergent symmetry group $\sfin$.
However, as we will see in the context of the subsystem codes, the symmetry group $\mathcal{S}_{\textrm{ini}}$ of the initial system may not only contain the $Z$-type initial symmetries from $\sini$.
Rather, $\mathcal{S}_{\textrm{ini}}$ may also be generated by some $X$-type initial symmetries forming a group $\mathcal{S}^X_{\textrm{ini}}$, i.e.,
\begin{equation}
\mathcal{S}_{\textrm{ini}} = \langle S^Z, S^X | S^Z \in \sini, S^X \in \mathcal{S}^X_{\textrm{ini}} \rangle.
\end{equation}
Note that the ungauging map $\widetilde\Gamma$ is still well-defined on the symmetric subspace $\mathcal{H}(\mathcal{S}_{\textrm{ini}})$, since $\mathcal{H}(\mathcal{S}_{\textrm{ini}})$ is contained in $\mathcal{H}(\sini)$.
The ungauging map $\widetilde\Gamma$ will transform the $X$-type initial symmetries $\mathcal{S}^X_{\textrm{ini}}$ into what we call the $X$-type preserved symmetries $\mathcal{S}^X_{\textrm{pre}}$. 
Thus, after ungauging of $Z$-type symmetries $\sini$ of the initial system, the symmetry group $\mathcal{S}_{\textrm{fin}}$ of the final system will not only be generated the emergent $X$-type symmetries $\sfin$ but also the preserved $X$-type symmetries $\mathcal{S}^X_{\textrm{pre}}$, i.e.,
\begin{equation}
\mathcal{S}_{\textrm{fin}} = \langle S_f^Z, S_p^X | S_f^Z \in \sfin, S_p^X \in \mathcal{S}^X_{\textrm{pre}} \rangle.
\end{equation}
We depict the relations between symmetries $\mathcal{S}_{\textrm{ini}}$  and $\mathcal{S}_{\textrm{fin}}$ of the initial and final systems in Fig.~\ref{fig_ungauging_symmetries}.

The fact that the symmetry group $\mathcal{S}_{\textrm{fin}}$ of the final system is generated by both the emergent and preserved symmetries will play a key role in the discussion of partial gauging in Sec.~\ref{sec_partial}.
Namely, in order to define the gauging map $\Gamma$ as the inverse of $\widetilde\Gamma$ we will be allowed to gauge only the $X$-type emergent symmetries $\sfin$, but not the $X$-type preserved symmetries $\mathcal{S}^X_{\textrm{fin}}$.

\subsection{Operational intepretation of ungauging}

The fact that both CSS stabilizer codes and the ungauging procedure can be characterized by chain complexes suggests a natural operational interpretation of gauging and ungauging through the lens of quantum error correction.
In the chain complex description of a CSS stabilizer code in Eq.~(\ref{eq_chain_CSS}), the boundary operator $\partial_X$ generates an error syndrome caused by a $Z$-type Pauli error.
In the ungauging chain complex in Eq.~(\ref{eq_chain_ungauge}), the boundary operator $\partial_X$ allows to define the ungauging map $\widetilde{\Gamma}$ via Eqs.~(\ref{eq_gaugeZ})~and~(\ref{eq_gaugeX}).
Thus, given a symmetric wavefunction $|\psi\rangle$ in $\mathcal{H}(\sini)$, the ungauging map $\widetilde{\Gamma}$ generates a configuration of error syndromes associated with $|\psi\rangle$ in the Pauli-$X$ basis (not in a computational basis).
Conversely, gauging can be interpreted as a procedure of obtaining a symmetric wavefunction $|\psi\rangle$ which is consistent with some error syndrome of the underlying CSS stabilizer code.

Since error syndromes are simply different configurations of excitations associated with violated stabilizer generators of the topological stabilizer code, the emergent symmetries $\mathcal{S}^X_{\text{fin}}$ can be interpreted as the law of conservation of excitations.
Namely, given a symmetric wavefunction $|\psi\rangle$ in $\mathcal{H}(\sini)$, the ungauging map $\widetilde{\Gamma}$ generates a configuration of excitations associated with the $X$-type stabilizer generators.
{As one can see from the next subsection,} violations of $X$-type stabilizer generators of the 2D toric code, which are often called electric charges, always appear in pairs and thus the total number of electric charges is conserved modulo two.
We will find that the emergent $X$-type symmetry $\sfin$ after the ungauging map $\widetilde{\Gamma}$ is a global $\mathbb{Z}_{2}$ symmetry generated by $\prod_{v} X_v$, which imposes the $\mathbb{Z}_{2}$ conservation law in the ungauged symmetric subspace $\mathcal{H}(\sfin)$ of the final system.

\subsection{Ungauging the toric code}

Now, we illustrate the preceding discussion by presenting the ungauging procedure for the 2D toric code.
For simplicity, we assume that the toric code is defined on a sphere, so that the ground state $\ket{\psi_{\textrm{TC}}}$ of the toric code Hamiltonian $H_{\textrm{TC}}$ is unique.
This assumption is equivalent to having no logical qubits, which in turn can be phrased as the CSS chain complex in Eq.~(\ref{eq_chain_CSS}) being exact, i.e., $\ker \partial_X = \im \partial_Z$.
We remark that ungauging for CSS stabilizer codes with a non-zero number of logical qubits can be achieved by redefining the CSS chain complex to include logical $Z$ operators into the $Z$-type stabilizers and adjusting the boundary map $\partial_Z $ appropriately, so that the resulting modified chain complex is exact

By definition, we want to ungauge $Z$-type symmetries of the toric code.
We choose the initial $Z$-symmetry group $\sini$ to be generated by $Z$-face stabilizers of the toric code,
$\sini = \langle Z(\partial_2 f) | f\in \face 2 {\mathcal{L}} \rangle$.
The symmetric Pauli group $\mathcal{P}(\sini)$ is then generated by single-qubit $Z_e$ operators and $X$-vertex stabilizers $X(\partial^T_1 v)$ of the toric code,
$\mathcal{P}(\sini) = \langle Z_e, X(\partial^T_1 v) | e\in \face 1 {\mathcal{L}}, v\in \face 0 {\mathcal{L}} \rangle$.
There is only one non-trivial relation between symmetric $X$-operators, namely
$\prod_{v\in\face 0 {\mathcal{L}}} X(\partial^T_1 v) = I$.
This leads us to the ungauging chain complex for the 2D toric code
\begin{equation}
\begin{tikzcd}[row sep = 0pt]
\raisebox{-8pt}{\stackanchor[8pt]{$C_2$}{$Z$-stabilizers}} \arrow[r,"\partial_2"] &
\raisebox{-8pt}{\stackanchor[8pt]{$C_1$}{qubits}} \arrow[r,"\partial_1"] &
\raisebox{-8pt}{\stackanchor[8pt]{$C_0$}{$X$-stabilizers}} \arrow[r,"(11\ldots1)"] &
\raisebox{-8pt}{\stackanchor[8pt]{$\mathbb{Z}_2$}{$X$-relations}}\\
\figbox{1.}{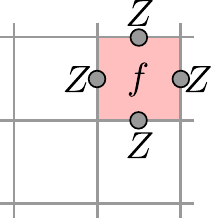} & \figbox{1.}{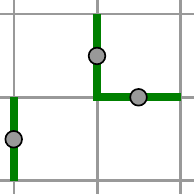} & \figbox{1.}{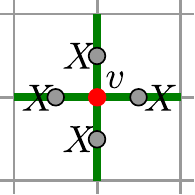} & \figbox{1}{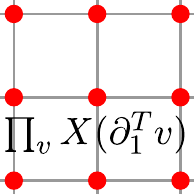}
\end{tikzcd}
\end{equation}
The ungauging map $\widetilde\Gamma$ transforms between subspaces $\hini$ and $\hfin$ of two Hilbert spaces $\mathcal{H}_{\textrm{ini}}$ and $\mathcal{H}_{\textrm{fin}}$, which are identified with qubits on edges and vertices of the lattice $\mathcal{L}$, respectively.
The emergent $X$-type symmetry group $\sfin$ is generated by only one operator,
$\sfin = \langle \prod_{v\in \face{0}{\mathcal{L}}} X_v \rangle$.
The symmetric operators from $\mathcal{P}(\sini)$ are mapped by $\widetilde\Gamma$ according to the prescription
\begin{eqnarray}
Z_e = \figbox{1}{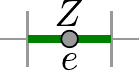} &\quad\xmapsto{\widetilde\Gamma}\quad& Z(\partial_1 e) = \figbox{1}{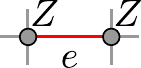},\\
X(\partial_1^T v) = \figbox{1}{eq/eq_tc3}  &\xmapsto{\widetilde\Gamma} & X_v = \figbox{1}{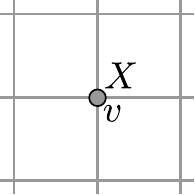},
\end{eqnarray}
and thus their commutation relations are preserved.
One can see that all initial $Z$-symmetry operators from $\sini$ are mapped to the identity operator.

We can easily see the ungauging map $\widetilde\Gamma$ transforms the toric code Hamiltonian $H_{\textrm{TC}} = -\sum_{v\in \face{0}{\mathcal{L}}} X(\partial_1^T v)$ into the paramagnet Hamiltonian $H_{\textrm{para}} = -\sum_{v \in \face{0}{\mathcal{L}}} X_v$, i.e.,
\begin{equation}
H_{\textrm{TC}} = -\sum_{v\in \face{0}{\mathcal{L}}} \figbox{1}{eq/eq_tc3} \ugmap 
H_{\textrm{para}} = -\sum_{v\in \face{0}{\mathcal{L}}} \figbox{1}{eq/eq_tc_Xb}
\end{equation}
Correspondingly, by ungauging the $Z$-symmetries of the unique ground state $\ket{\psi_{\textrm{TC}}}$ of the toric code we obtain the unique ground state $\ket{\psi_{\textrm{para}}} = \ket{+}^{\otimes |\face 0 {\mathcal{L}}|}$ of the paramagnet.
We emphasize that $\ket{\psi_{\textrm{TC}}}$ satisfies local $\mathbb{Z}_2$ symmetries from the $Z$-type symmetry group $\mathcal{S}_Z$, whereas $\ket{\psi_{\textrm{para}}}$ has a global $\mathbb{Z}_2$ symmetry generated by $\prod_{v\in \face{0}{\mathcal{L}}} X_v$.

We remark that in a similar way we can ungauge higher-form symmetries.
For instance, consider the $D$-dimensional toric code of type $k\in \{ 1, \ldots, D-1\}$, where $k$ indicates that qubits are placed on $k$-dimensional cells of the lattice $\mathcal{L}$.
For simplicity, we assume that the lattice $\mathcal{L}$ is on a $D$-dimensional sphere, so that the toric code has no logical qubits.
If we choose to ungauge $Z$-type symmetries of the toric code, which are associated with the $(k+1)$-dimensional cells of $\mathcal{L}$, then the symmetric $X$-operators are identified with $(k-1)$-cells.
There is only one relation between symmetric $X$-operators, i.e., the product of all of them is the identity operator.
Thus, the ungauging chain complex for the $D$-dimensional toric code of type $k$ is
$C_{k+1} \xrightarrow{\partial_{k+1}} C_k \xrightarrow{\partial_k} C_{k-1} \xrightarrow{(11\ldots 1)} \mathbb{Z}_2$
and we can show that the toric code Hamiltonian is mapped by $\widetilde\Gamma$ into the paramagnet Hamiltonian with spins on $(k-1)$-dimensional cells and a global $\mathbb{Z}_2$ symmetry.

\subsection{Ungauging the Bacon-Shor code}

Before we start talking about ungauging the Bacon-Shor code, we briefly discuss subsystem codes, which are a generalization of stabilizer codes.
A subsystem code is specified by the gauge group $\mathcal{G}_{\textrm{sub}}$, which is a subgroup of the Pauli group $\mathcal{P}$.
Unlike in the stabilizer case, we do not require that the gauge group $\mathcal{G}_{\textrm{sub}}$ be Abelian.
The stabilizer group $\mathcal{S}_{\textrm{sub}}$ of the subsystem code can be found as the center of the gauge group in the Pauli group $\mathcal{P}$.
Representatives of non-trivial bare logical operators are the elements of the centralizer of the gauge group $\mathcal{G}_{\textrm{sub}}$ which do not belong to $\mathcal{G}_{\textrm{sub}}$.
For a CSS subsystem code the generators of $\mathcal{G}_{\textrm{sub}}$ can be chosen to be of $X$-type or $Z$-type.

\begin{figure}[h!]
\centering
\includegraphics[width=.65\columnwidth]{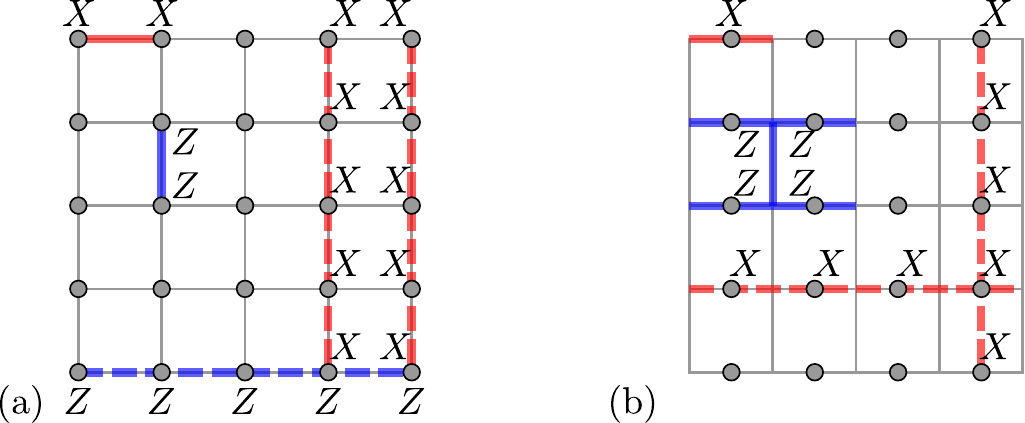}
\caption{
(a) The subsystem Bacon-Shor code with qubits placed on vertices of a square lattice on a torus.
The $X$- and $Z$-type gauge generators are identified with horizontal $e_H$ (red) and vertical $e_V$ (blue) edges, respectively.
We depict an $X$-type stabilizer (dashed red) and a representative of the bare logical $\overline Z$ operator (dashed blue).
(b) The Xu-Moore model with qubits on horizontal edges.
The Hamiltonian $H_{XM}$ of the model is a sum of single-qubit $X$ and four-qubit $Z$ (blue) operators.
The symmetries of the model are $X$-type horizontal and vertical operators (dashed red).
}
\label{fig_BS} 
\end{figure}

The 2D Bacon-Shor code is an example of a CSS subsystem code, which is defined by placing qubits on the vertices of a square lattice on a torus.
The gauge group $\mathcal{G}_{\textrm{BS}}$ is generated by operators associated with edges of the lattice.
Namely, for every edge, either horizontal $e_H$ and vertical $e_V$, there is a two-qubit gauge generator, respectively of $X$-type or $Z$-type.
The stabilizer group $\mathcal{S}_{\textrm{BS}}$ is generated by two-column $X$-type and two-row $Z$-type operators.
The Bacon-Shor code has one logical qubit and the bare logical $\overline X$ and $\overline Z$ operators can be represented as a single-column $X$-type and a single-row $Z$-type operators, respectively.
For an illustration of the 2D Bacon-Shor code, see Fig.~\ref{fig_BS}(a).

Now we are going to discuss the ungauging procedure for the Bacon-Shor code.
The initial symmetry group $\mathcal{S}_{\textrm{ini}}$ of the Bacon-Shor code contains all the stabilizers from $\mathcal{S}_{\textrm{BS}}$.
Since the code has one logical qubit, we also include the representatives of the bare logical $\overline Z$ in $\mathcal{S}_{\textrm{ini}}$.
We want to ungauge all $Z$-type symmetries of the Bacon-Shor model, which include $Z$-type stabilizers and bare logical $\overline Z$.
Note that any $Z$-type stabilizer can be written as a product of two representatives of $\overline Z$, and thus the initial $Z$-type symmetry group $\sini$ is generated by representatives of $\overline Z$.
One can verify that the symmetric Pauli subgroup $\mathcal{P}(\sini)$ is generated by single-qubit $Z$ operators, as well as by the $X$-type gauge generators.
Let us label rows of the lattice by the elements $l \in \mathcal{B}_R$ and denote by $\textrm{row}(l)$ the set of horizontal edges belonging to the row $l$.
Note that for every row $l$ there is one relation, namely the product of $X$-type gauge generators associated with all edges from $\textrm{row}(l)$ is the identity operator.
Thus, we arrive at the ungauging chain complex for the Bacon-Shor code
\begin{equation}
\begin{tikzcd}[row sep = 0pt]
\raisebox{-8pt}{\stackanchor[8pt]{$C_{\textrm{row}}$}{$Z$-bare logical}} \arrow[r,"\partial_Z"] &
\raisebox{-8pt}{\stackanchor[8pt]{$C_{\textrm{vertex}}$}{initial qubits}} \arrow[r,"\partial_X"] &
\raisebox{-8pt}{\stackanchor[8pt]{$C_{\textrm{h-edge}}$}{$X$-gauge}} \arrow[r,"\partial_R"] &
\raisebox{-8pt}{\stackanchor[8pt]{$C_{\textrm{row}}$}{$X$-relations}}\\
\figbox{1}{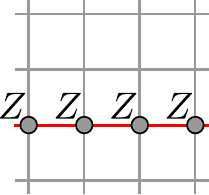} & \figbox{1}{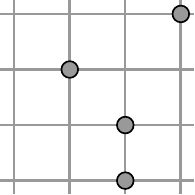} & \figbox{1}{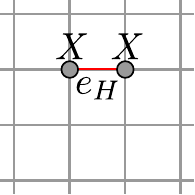} & \figbox{1}{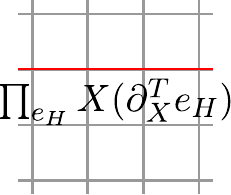}
\end{tikzcd}
\end{equation}
From the above ungauging chain complex we can infer that the final Hilbert space $\mathcal{H}_{\textrm{fin}}$ is identified with qubits placed on horizontal edges of the lattice.
Moreover, the subspace $\hfin$ of $\mathcal{H}_{\textrm{fin}}$ is determined by the emergent $X$-type symmetry group $\sfin$, which is generated by $X$-type operators associated with every row $l$, i.e.,
$\sfin = \langle \prod_{e_H\in \textrm{row}(l)} X_{e_H} | \forall l \in \mathcal{B}_R \rangle$.
Lastly, the symmetric operators from $\mathcal{P}(\sini)$ are mapped by the ungauging map $\widetilde\Gamma$ according to the prescription
\begin{eqnarray}
\label{eq_ungaguge_BS1}
Z_v = \figbox{1.}{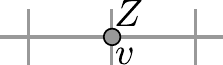} &\quad\xmapsto{\widetilde\Gamma}\quad& Z(\partial_X v) = \figbox{1.}{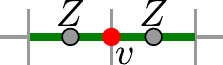},\\
\label{eq_ungaguge_BS2}
X(\partial_X^T e_H) = \figbox{1.}{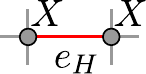} &\xmapsto{\widetilde\Gamma} & X_{e_H} = \figbox{1.}{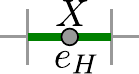}.
\end{eqnarray}

Now, let us consider the following Hamiltonian for the Bacon-Shor code
\begin{equation}
H_{\textrm{BS}} = J_X H^X_{\textrm{BS}} + J_Z H^Z_{\textrm{BS}}
= - J_{X} \sum_{e_H} \figbox{1.}{eq/eq_bs2a} - J_{Z} \sum_{e_V} \figbox{1.}{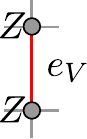}.
\end{equation}
Ground states of $H_{\textrm{BS}}$ satisfy stabilizer symmetries with $+1$ eigenvalues.
In fact, a similar statement can be proven for arbitrary CSS subsystem Hamiltonians with negative coefficients via the Perron-Frobenius theorem, see~\cite{Ocko11} for instance.
As we have already mentioned, the symmetry group of the Bacon-Shor model $\mathcal{S}_{\textrm{ini}}$ is generated by two types of operators, the $Z$-type representatives of the bare logical $\overline Z$ and the $X$-type stabilizers from $\mathcal{S}_{\textrm{BS}}$.
Using Eqs.~(\ref{eq_ungaguge_BS1})~and~(\ref{eq_ungaguge_BS2}), we find that the procedure of ungauging of the initial $Z$-symmetry group $\sini$ transforms the Bacon-Shor Hamiltonian $H_{\textrm{BS}}$ into the following
\begin{equation}
H_{\textrm{XM}} = J_X H^X_{\textrm{XM}} + J_Z H^Z_{\textrm{XM}}
= - J_{X} \sum_{e_H} \figbox{1.}{eq/eq_bs2b} - J_{Z} \sum_{e_V} \figbox{1.}{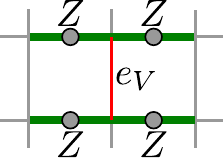},
\end{equation}
which describes the so-called Xu-Moore model~\cite{Xu04}.
We emphasize that the final symmetry group $\mathcal{S}_{\textrm{fin}}$ of the model with the Hamiltonian $H_{\textrm{XM}}$ is generated by the emergent $X$-type symmetries from $\sfin$ and the preserved $X$-type symmetries from $\mathcal{S}^X_{\textrm{pre}}$.
The emergent and preserved symmetries correspond to respectively horizontal and vertical $X$-type operators; see Fig.~\ref{fig_BS}(b).

We remark that the recent independent work~\cite{You18} has focused on SPT phases protected by sub-manifold symmetries, similar to those in the ungauged Bacon-Shor code.
Our result on the Bacon-Shor code not only hints at a natural gauging procedure for sub-manifold SPT phases, but also suggests a further generalization by using the language of CSS subsystem codes.
This question can be partly answered by our framework of creating SPT Hamiltonians protected by unconventional symmetries, which we present in Sec.~\ref{sec_fracSPT}.

\subsection{Partial gauging of symmetries}
\label{sec_partial}

We remark that instead of ungauging we can consider a reverse procedure, related to the standard gauging procedure.
For instance, we could start with the ground state $\ket{\psi_{\textrm{para}}}$ of the paramagnet and by gauging its global $\mathbb{Z}_2$-symmetry we would obtain the toric code ground state $\ket{\psi_{\textrm{TC}}}$ with a local $\mathbb{Z}_2$-symmetry.
The gauging map $\Gamma$ can be operationally defined in the same way as the ungauging map $\widetilde\Gamma$ by replacing $X$ and $Z$ in the discussion in Sec.~\ref{sec_definition}.
Importantly, this definition leads to the gauging map $\Gamma$, which is the inverse of the ungauging map $\widetilde\Gamma$.
This is in agreement with the fact that gauging and ungauging maps should be invertible onto the symmetric subspaces $\hini$ and $\hfin$.

Let us revisit the example from the previous subsection and try proceeding in the other direction, i.e., obtain the Bacon-Shor model by gauging $X$-type symmetries of the Xu-Moore model. 
However, if we gauged all $X$-type symmetries of the Xu-Moore model, we would not get the Bacon-Shor model!
One can verify that choosing to gauge the symmetry group $\mathcal{S}_{\textrm{fin}}$ generated by the $X$-type horizonal and vertical operators would lead to the following transformation of symmetric Pauli operators under the gauge map $\Gamma$
\begin{eqnarray}
\figbox{1.}{eq/eq_xm_Z} \ugmap \figbox{1.}{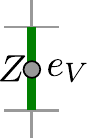},\qquad
\figbox{1.}{eq/eq_bs2b} \ugmap \figbox{1.}{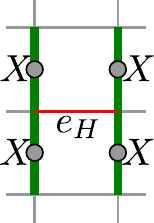}
\end{eqnarray}
Thus, the Hamiltonian $H_{\textrm{XM}}$ of the Xu-Moore model would be mapped to $H'_{\textrm{XM}}$, which is the same as $H_{\textrm{XM}}$ but with qubits placed on vertical edges (instead of horizontal) and $X$ and $Z$ operators swapped, i.e., $\tilde H_{\textrm{XM}} \simeq \overline{H} (H_{\textrm{XM}}) \overline{H}^\dag$, where $\overline{H}$ denotes the transversal Hadamard gate applied to all the qubits.

In order to map the Xu-Moore model back into the Bacon-Shor code, we need to gauge only some of its $X$-type symmetries.
In particular, if we choose to gauge the emergent $X$-symmetry group $\mathcal{S}^X_{\textrm{fin}}$ generated by $X$-type horizonal operators, then the symmetric Pauli group is given by
\begin{equation}
\mathcal{P}(\mathcal{S}^X_{\textrm{fin}}) =
\left\langle \figbox{1.}{eq/eq_bs2b}, \figbox{1.}{eq/eq_bs1b} \middle| \forall e_H, v \right\rangle,
\end{equation}
and the gauge map $\Gamma$ would transform symmetric Pauli operators from $\mathcal{P}(\mathcal{S}^X_{\textrm{fin}})$ according to the prescription
\begin{eqnarray}
\label{eq_gauge_XM1}
Z(\partial_X v) = \figbox{1.}{eq/eq_bs1b} &\quad\xmapsto{\Gamma}\quad& Z(v) = \figbox{1.}{eq/eq_bs1a},\\
\label{eq_gauge_XM2}
X(e_H) = \figbox{1.}{eq/eq_bs2b} &\xmapsto{\Gamma} & X(\partial_X^T e_H) = \figbox{1.}{eq/eq_bs2a}.
\end{eqnarray}
We easily check that the Hamiltonian $H_{\textrm{XM}}$ of the Xu-Moore model is mapped by $\Gamma$ into the Hamiltonian $H_{\textrm{BS}}$ of the Bacon-Shor model and the initial symmetries from $\mathcal{S}_{\textrm{ini}}$ of the Bacon-Shor model are recovered.

We remark that the Xu-Moore model~\cite{Xu04} which undergoes the first order phase transition.
Namely, when one changes the relative strength of $J_{X}$ and $J_{Z}$, then there is a phase transition for $J_X = J_Z$~\cite{Orus09}.
The presence of a phase transition suggests a possibility of $H_{\textrm{XM}}^{X}$ and $H_{\textrm{XM}}^{Z}$ being in different SPT phases of matter when the $X$-type horizontal and vertical symmetries from $\mathcal{S}_{\textrm{fin}}$ are enforced.
This intuition can be made rigorous by considering the gauging map $\Gamma$ specified by Eqs.~(\ref{eq_gauge_XM1})~and~(\ref{eq_gauge_XM2}).
When $J_{X}\gg J_{Z}$ (respectively, $J_{Z}\gg J_{X}$), a ground state exhibits a global entanglement structure similar to one of the GHZ states extending in the horizontal (vertical) direction, implying that two phases cannot be connected by local unitary transformations.
Another way of arguing the phase separation is to note that the geometric shapes of logical operators are topologically distinct, and thus the Hamiltonians $H_{\textrm{BS}}^{X}$ and $H_{\textrm{BS}}^{Z}$ have to represent two different SPT phases of matter under the Bacon-Shor stabilizer symmetries $\mathcal{S}_{\textrm{BS}}$.
At the same time, $H_{\textrm{BS}}^{X}$ and $H_{\textrm{BS}}^{Z}$ can be obtained from $H_{\textrm{XM}}^{X}$ into $H_{\textrm{XM}}^{Z}$ by gauging the $X$-type symmetry group $\sfin$.
This in turn shows that there is no short-depth circuit respecting the symmetry group $\mathcal{S}_{\textrm{fin}}$, which transforms $H_{\textrm{XM}}^{X}$ into $H_{\textrm{XM}}^{Z}$ without closing the gap.
We conclude that $H_{\textrm{XM}}^{X}$ into $H_{\textrm{XM}}^{Z}$ belong to different SPT phases.

\section{Ungauging the gauge color code}
\label{sec_gcc}

In this section, we look at the 3D gauge color code, which is another example of a CSS subsystem code.
The 3D gauge color code exhibits advantages for fault-tolerant quantum computing, such as logical operations with relatively low overhead.
Moreover, the 3D gauge color code allows one to reliably detect measurement errors in a single time step. 
Here, we apply the procedure of ungauging from Sec.~\ref{sec_definition} to the 3D gauge color code and see that the ungauged model corresponds to six (decoupled) copies of $\mathbb{Z}_2$ lattice gauge theory.
We also discuss various quantum phases of matter arising from different Hamiltonians associated with the gauge color code.
In particular, we show that depending on the choice of the initial Hamiltonian the procedure of ungauging $Z$-type stabilizer symmetries leads to the paramagnet, the 3D toric code, and the RBH model, which represent distinct SPT phases of matter in the presence of the stabilizer symmetries of the GCC.

\subsection{3D gauge color code}
\label{sec_GCC_def}

A standard way of defining the 3D GCC is to place qubits on the vertices of a three-dimensional lattice, which is $4$-valent and its volumes are $4$-colorable; see Ref.~\cite{Kubica15b} for details.
The gauge group of the 3D GCC is generated by the $X$- and $Z$-type operators supported on qubits associated with every face of the lattice.
One can show that the stabilizer group of the 3D GCC is generated by volume operators of $X$- and $Z$-type.
However, we will present an equivalent definition of the 3D GCC on the (dual) lattice built of tetrahedra, where qubits are placed on tetrahedra, whereas gauge and stabilizer generators are identified with edges and and vertices of the (dual) lattice.
We choose to use the dual lattice since this simplifies the discussion of lattice gauge theory and stabilizer Hamiltonians associated with the 3D GCC.

\begin{figure}[h!]
\centering
\includegraphics{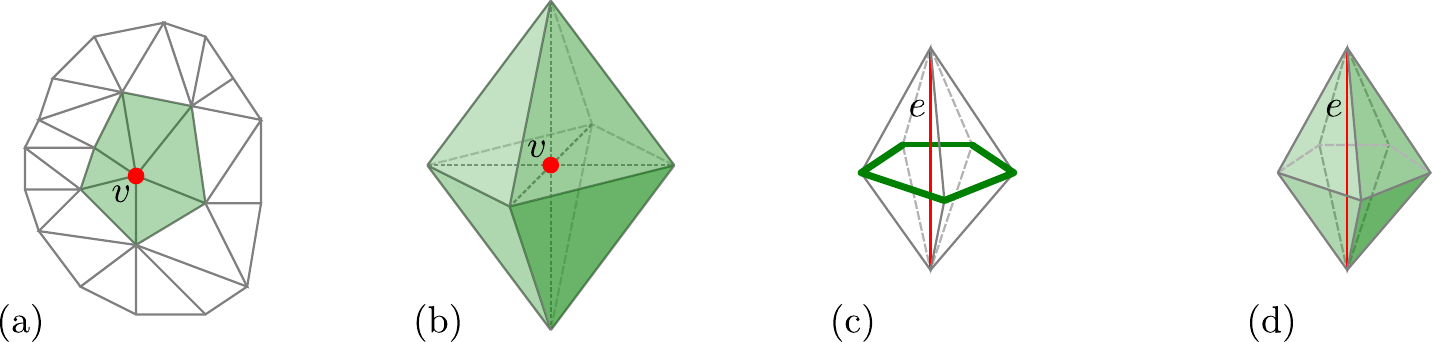}
\caption{
Examples of stars and links in 2D and 3D.
In (a) and (b), we illustrate the $2$-star $\star 2 v$ and the $3$-star $\star 3 v$ of the vertex $v$ (red), which respectively are the sets of six triangular faces and eight tetrahedra (shaded in green) containing $v$.
(c) The $1$-link $\link 1 e$ of the edge $e$ (red) is the set of five edges (green), each of which belongs to the same tetrahedron as $e$ but does not overlap with $e$.
(d) The $3$-star $\star 3 e$ of the edge $e$ (red) is the set of five tetrahedra (shaded in green) containing $e$.
}
\label{fig_star_link} 
\end{figure}

In order to discuss the color code, we need to introduce a couple of notions from combinatorial geometry.
Let $\mathcal{L}$ be a $D$-dimensional lattice
\footnote{More specifically, we assume that $\mathcal{L}$ is a homogeneous simplicial $D$-complex containing a finite number of simplices.}
built of $D$-simplices, where $D\geq 2$.
We denote by $\face i {\mathcal{L}}$ the set of all $i$-simplices of $\mathcal{L}$, where $i\in \{0,1,\ldots, D\}$.
In particular, $\face 0 {\mathcal{L}}$, $\face 1 {\mathcal{L}}$ and $\face 2 {\mathcal{L}}$ correspond to the sets of vertices, edges and triangular faces of $\mathcal{L}$, respectively.
Similarly, we denote by $\face i \delta$ the set of all $i$-simplices contained in a $k$-simplex $\delta\in \face k {\mathcal{L}}$. 
We define the $n$-star $\star n \delta$ of  the simplex $\delta$ to be the set of all $n$-simplices containing $\delta$, namely
\begin{equation}
\star n \delta = \{ \epsilon \in \face n {\mathcal{L}} |  \delta \subset \epsilon \}.
\end{equation}
We also need the notion of the $n$-link $\link n \delta$, which is the set of all $n$-simplices, which do not intersect $\delta$ but at the same time belong to the same $D$-simplex as $\delta$, namely
\begin{equation}
\link n \delta = \{ \epsilon \in \face n {\mathcal{L}} | \epsilon \cap \delta = \emptyset \wedge \exists \sigma\in\face D {\mathcal{L}} : \delta, \epsilon \subset \sigma \}.
\end{equation}
We illustrate the notions of the star and the link with examples in two and three dimensions in Fig.~\ref{fig_star_link}. 
Lastly, we say that $\mathcal{L}$ is $(d+1)$-colorable, if we can assign different $d+1$ colors to vertices $\face 0 {\mathcal{L}}$ in such a way that no two vertices incident to the same edge have the same color.
We will denote by $\col \delta$ the set of colors of vertices which belong to the simplex $\delta$.

To facilitate the discussion we need to introduce a generalized boundary operator $\bnd k l : C_k \rightarrow C_l$ for $k\neq l$.
The generalized boundary operator $\bnd k l$  is a linear map between two $\mathbb{F}_2$-vector spaces $C_k$ and $C_l$ with bases $\face k {\mathcal{L}}$ and $\face l {\mathcal{L}}$, respectively.
The map $\bnd k l$ is defined by specifying its action on every basis element $\delta\in\face k {\mathcal{L}}$, namely
\begin{equation}
\bnd k l \delta =
\begin{cases}
\sum_{\sigma \in \face l \delta} \sigma \quad\textrm{ if } k>l,\\
\sum_{\sigma \in \star l \delta} \sigma \quad\textrm{ if } k<l.
\end{cases}
\end{equation}
We remark that: $(i)$ in this notation the standard boundary operator $\partial_i = \bnd i {i-1}$, $(ii)$ in general, the composition of two generalized boundary operators is not the zero operator, i.e., $\bnd l m \circ \bnd k l \neq 0$, and $(iii)$ if one views the generalized boundary operators as binary matrices, then $(\bnd k l)^T = \bnd l k$.

Now, we describe the 3D gauge color code defined on the 3D lattice $\mathcal{L}$ without boundaries.
We assume that $\mathcal{L}$ consists of tetrahedra and is $4$-colorable.
We place one qubit on every tetrahedron of the lattice $\mathcal{L}$.
The gauge group $\mathcal{G}_{\textrm{GCC}}$ of the 3D gauge color code is generated by the $X$- and $Z$-type operators identified with the edges of the lattice $\mathcal{L}$, namely
\begin{eqnarray}
\mathcal{G}_{\textrm{GCC}} &=& \langle X(\bnd 1 3 e), Z(\bnd 1 3 e) | e \in \face 1 {\mathcal{L}} \rangle
= \left\langle X\left( \figbox{1}{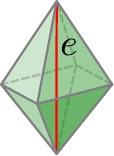} \right), Z\left( \figbox{1}{eq/eq_star_e} \right) \right\rangle.
\label{eq_gcc_gauge}
\end{eqnarray}
Note that in Eq.~(\ref{eq_gcc_gauge}) we use the notation $X(\cdot)$ and $Z(\cdot)$ to represent the $X$-type and $Z$-type operators supported on all the qubits identified with tetrahedra depicted in the schematic; we will use this pictorial notation in the rest of the paper.
In other words, a gauge generator identified with an edge $e$ is supported on qubits identified with the tetrahedra in the neighborhood of the edge $e$.
One finds that the stabilizer group $\mathcal{S}_{\textrm{GCC}}$ of the gauge color code is generated by $X$- and $Z$-type operators associated with the vertices of $\mathcal{L}$, i.e.,
\begin{eqnarray}
\mathcal{S}_{\textrm{GCC}} &=& \langle X(\bnd 0 3 v), Z(\bnd 0 3 v) | v \in \face 0 {\mathcal{L}} \rangle
= \left\langle X\left( \figbox{1}{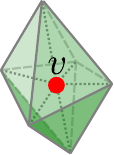} \right), Z\left( \figbox{1}{eq/eq_star_v} \right) \right\rangle.
\end{eqnarray}
In other words, a stabilizer generator identified with a vertex $v$ is supported on qubits on all neighboring tetrahedra, which contain the vertex $v$.

\subsection{Ungauging the gauge color code}

We want to ungauge the initial $Z$-type symmetry group $\sini$ of the gauge color code, which is generated by $Z$-type stabilizers from $\mathcal{S}_{\textrm{GCC}}$.
Since we consider the lattice $\mathcal{L}$ without boundaries, the gauge color code has zero logical qubits and thus we do not need to include bare logical $\overline Z$ operators in $\sini$.
One can check that the symmetric Pauli subgroup $\mathcal{P}(\sini)$ is generated by single-qubit $Z$ operators and the $X$-type gauge generators, i.e.,
$\mathcal{P}(\sini) = \langle Z_t, X(\bnd 1 3 e) | \forall t\in \face 3 {\mathcal{L}}, e \in \face 1 {\mathcal{L}}\rangle$.

Now we discuss relations between the $X$-type gauge generators.
For convenience, we label four different colors assigned to vertices by $a$, $b$, $c$ and $d$. 
Let us consider a vertex $v$ of color $a$.
Note that the $X$-type stabilizer associated with $v$ is given by $X(\bnd 0 3 v)$.
At the same time, we can express $X(\bnd 0 3 v)$ as the product of $X$-type gauge generators associated with edges of color $ab$ (similarly for edges of color $ac$ or $ad$), which are incident to $v$. Namely,
\begin{alignat}{6}
X(\bnd 0 3 v) &=& \prod_{\substack{e\in \bnd 0 1 v \\ \col{e} = ab}} X(\bnd 1 3 e) &=& 
\prod_{\substack{e\in \bnd 0 1 v \\ \col{e} = ac}} X(\bnd 1 3 e) &=& \prod_{\substack{e\in \bnd 0 1 v \\ \col{e} = ad}} X(\bnd 1 3 e).
\end{alignat}
We conclude that by multiplying the $X$-type gauge generators identified with edges incident to $v$, whose color is in $\mathbb{C} = \{ ab, ac\}$ (similarly for two other pairs of colors, $\mathbb{C} = \{ ab, ad \}$ and  $\mathbb{C} = \{ ac, ad \}$) we get the identity operator.
Thus, for every vertex $v\in\face 0 {\mathcal{L}}$ there are three relations for the $X$-type gauge generators
\begin{equation}
\forall v\in \face 0 {\mathcal{L}} : \prod_{\substack{e\in \bnd 0 1 v \\ \col{e} \in\mathbb{C} }} X(\bnd 1 3 e) = I,
\label{eq_gcc_relations}
\end{equation}
since there are ${3 \choose 2} = 3$ ways of picking $\mathbb{C} \subset \{ ab, ac, ad\}$.
At the same time, only two relations are independent.

We are ready to specify the ungauging chain complex for the gauge color code
\begin{equation}
\begin{tikzcd}[row sep = 0pt]
\raisebox{-8pt}{\stackanchor[8pt]{$C_0$}{$Z$-stabilizers}} \arrow[r,"\bnd 0 3"] &
\raisebox{-8pt}{\stackanchor[8pt]{$C_3$}{qubits}} \arrow[r,"\bnd 3 1"] &
\raisebox{-8pt}{\stackanchor[8pt]{$C_1$}{$X$-gauge}} \arrow[r,"\partial_R"] &
\raisebox{-8pt}{\stackanchor[8pt]{$C_R$}{$X$-relations}}\\
\figbox{1}{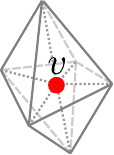} & \figbox{1}{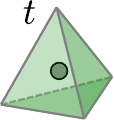} &
\figbox{1}{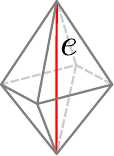} & \quad\figbox{1}{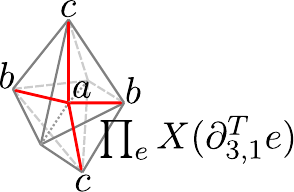}
\end{tikzcd}
\label{eq_chain_GCC}
\end{equation}
Note that in order to write the above ungauging chain complex in a way consistent with Eq.~(\ref{eq_chain_ungauge}) we use a fact that 
$\bnd 3 1 = (\bnd 1 3)^T$ and define $\partial_R$ according to Eq.~(\ref{eq_gcc_relations}).
The Hilbert space $\mathcal{H}_{\textrm{fin}}$ of the final ungauged model is associated with qubits placed on edges, i.e., $Q_{\textrm{fin}} = \face 1 {\mathcal{L}}$.
The subspace $\hfin$ of $\mathcal{H}_{\textrm{fin}}$ is determined by the emergent $X$-type symmetry group
\begin{eqnarray}
\sfin &=& \left\langle \prod_{\substack{e\in \bnd 0 1 v \\ \col{e} \in \mathbb{C} }} X_e \middle|
\forall v\in\face 0 {\mathcal{L}}, \mathbb{C} \subset \{ ab, ac, ad \} \wedge |\mathbb{C} | = 2 \right\rangle\\
&=& \left\langle X\left(\figbox{1}{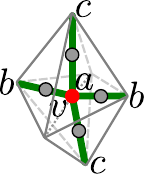}\right), X\left(\figbox{1}{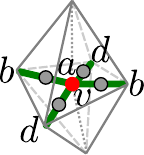}\right),
X\left(\figbox{1}{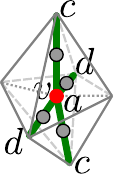}\right) \right\rangle.
\end{eqnarray}

The generators of the initial symmetric Pauli subgroup $\mathcal{P}(\sini)$ are mapped by the ungauging map $\widetilde\Gamma$ in the following way
\begin{eqnarray}
\label{eq_ungauge_GCC1}
\forall e\in \face 1 {\mathcal{L}}: X(\bnd 3 1 ^T e) = 
X\left( \figbox{1}{eq/eq_star_e} \right) & \ugmap& X_e = \figbox{1}{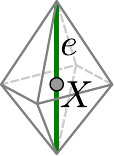},\\
\label{eq_ungauge_GCC2}
\forall t\in \face 3 {\mathcal{L}}: Z_t = \figbox{1}{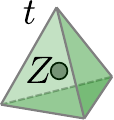} &\ugmap& Z(\bnd 3 1 t) = Z\left( \figbox{1}{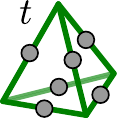} \right).
\end{eqnarray}

Let us now discuss how the Hamiltonian of the gauge color code $H_{\textrm{GCC}}$  transforms under the ungauging map $\widetilde\Gamma$, where we choose
\begin{equation}
H_{\textrm{GCC}} = -J_X\sum_{e \in \face 1 {\mathcal{L}}} X(\bnd 1 3 e) -J_Z\sum_{e \in \face 1 {\mathcal{L}}} Z(\bnd 1 3 e)
\end{equation}
to be the sum of all $X$- and $Z$-type gauge generators.
The initial symmety group $\mathcal{S}_{\textrm{ini}}$ of the model is the stabilizer group $\mathcal{S}_{GCC}$ and we ungauge only the $Z$-type stabilizers.
Using Eqs.~(\ref{eq_ungauge_GCC1})~and~(\ref{eq_ungauge_GCC2}) we find that the gauge generators transform under $\widetilde\Gamma$ in the following way
\begin{eqnarray}
X(\bnd 1 3 e) = X\left( \figbox{1}{eq/eq_star_e} \right) &\ugmap& X_e = \figbox{1}{eq/eq_gcc_Xb},\\
Z(\bnd 1 3 e) = Z\left(\figbox{1}{eq/eq_star_e} \right) &\ugmap& Z(\link 1 e) = Z\left( \figbox{1}{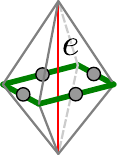}\right).
\end{eqnarray}
We can easily verify
\footnote{Namely, two operators $X(\bnd 1 3 {e_1})$ and $Z(\bnd 1 3 {e_2})$ anticommute iff edges $e_1$ and $e_2$ belong to some tetrahedron $t\in\face 3 {\mathcal{L}}$ but do not overlap, i.e., $e_1,e_2 \in \face 1 t$ and $e_1 \cap e_2 = \emptyset$.
The last condition can be restated as $e_1 \in \link 1 {e_2}$, which in turn is equivalent to $X_{e_1}$ and $Z(\link 1 {e_2})$ anticommuting.}
that the commutation relations of the operators in $\mathcal{G}_{GCC}$ are preserved by the ungauging map $\widetilde\Gamma$.
Thus, the Hamiltonian $H_{\textrm{GCC}}$ is mapped by $\widetilde\Gamma$ as follows
\begin{eqnarray}
H_{\textrm{GCC}} \ugmap H_{\textrm{LGT}}
&=& -J_X\sum_{e \in \face 1 {\mathcal{L}}} \figbox{1}{eq/eq_gcc_Xb}
- J_Z\sum_{e \in \face 1 {\mathcal{L}}} Z\left( \figbox{1}{eq/eq_link_e}\right).
\end{eqnarray}
Moreover, the final symmetry group $\mathcal{S}_{\textrm{fin}}$ of the model is generated by the $X$-type emergent and preserved symmetries, namely
\begin{eqnarray}
\mathcal{S}_{\textrm{fin}}
&=& \left\langle \prod_{\substack{e\in \bnd 0 1 v \\ \col{e} \in \{ ab, ac, ad \} }} X_e \middle| \forall v\in\face 0 {\mathcal{L}}\right\rangle \\
&=& \left\langle X\left(\figbox{1}{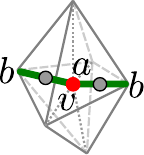}\right), X\left(\figbox{1}{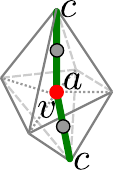}\right),
X\left(\figbox{1}{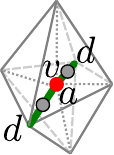}\right) \right\rangle 
\label{eq_LGT_fin}
\end{eqnarray}
We identify the model described by the Hamiltonian $H_{\textrm{LGT}}$ with the symmetry group $\mathcal{S}_{\textrm{fin}}$ as six decoupled copies of $\mathbb{Z}_2$ lattice gauge theory,
one copy for each pair of colors $ab$ (note that a pair of colors $ab$ can be chosen in ${4 \choose 2} = 6$ different ways).
The $ab$ copy is defined on a sublattice $\mathcal{L}^{ab}$ obtained from $\mathcal{L}$ by keeping only vertices of color $a$ or $b$ and edges between them; see Appendix in Ref.~\cite{Kubica15b} for a detailed discussion.
The qubits are placed on edges of color $ab$.
The $ab$ copy of lattice gauge theory respects a local $\mathbb{Z}_2$ symmetry group $\mathcal{S}_{\textrm{LGT}}^{ab}$, which is locally generated by  the $X$-type vertex operators
\begin{equation}
\mathcal{S}_{\textrm{LGT}}^{ab} = \langle X(\bnd 0 1 ^{ab} v) | v \in \face 0 {\mathcal{L}^{ab}} \rangle
= \left\langle X\left(\figbox{1}{eq/eq_gcc_Xnew1} \right)\right\rangle,
\label{eq_lgt_symab}
\end{equation}
where $\bnd 0 1 ^{ab}$ is the boundary operator restricted to the sublattice $\mathcal{L}^{ab}$.
Note that each copy of lattice gauge theory has similar $1$-form symmetries as the ones depicted in Eq.~(\ref{eq_lgt_symab}).
We emphasize that six copies of lattice gauge theory do not interact with one another, since they are defined on sublattices with different edges, and thus the sets of qubits for each copy are mutually disjoint.

We remark that in order to obtain the gauge color code Hamiltonian $H_{\textrm{GCC}}$ from $H_{\textrm{LGT}}$ we would need to partially gauge some $X$-type symmetries of lattice gauge theory, i.e., the $X$-type preserved symmetry group $\mathcal{S}_{\textrm{pre}}^X$.
Note that $\mathcal{S}_{\textrm{pre}}^X$ is generated by local products of $1$-form symmetries from any two copies of lattice gauge theory.
In particular, for each vertex $v$ of color $a$, there are three generators of $\mathcal{S}_{\textrm{pre}}^X$ associated with $v$, i.e.,
$X(\bnd 0 1 ^{ab} v)X(\bnd 0 1 ^{ac} v)$, $X(\bnd 0 1 ^{ab} v)X(\bnd 0 1 ^{ad} v)$ and $X(\bnd 0 1 ^{ac} v)X(\bnd 0 1 ^{ad} v)$.
In contrast, if we tried to gauge all the $X$-type symmetries from $\mathcal{S}_{\textrm{fin}}$, then we would obtain a model
$\tilde{H}_{\textrm{LGT}} = \overline H(H_{\textrm{LGT}})\overline H^\dag$ with the symmetries 
$\tilde{\mathcal{S}}_{\textrm{fin}} = \overline H(\mathcal{S}_{\textrm{fin}})\overline H^\dag$,
which is equivalent to lattice gauge theory $H_{\textrm{LGT}}$ and its symmetries $\mathcal{S}_{\textrm{fin}}$ via the transversal Hadamard gate $\overline H$ applied to all qubits.

\subsection{Quantum phases of the 3D gauge color code}

Given a subsystem code with the gauge group $\mathcal{G}_{\textrm{sub}}$, such as the 3D gauge color code, there are many possible ways of associating a physical model with that code.
Namely, we can adjust the couplings $J_i$ in the Hamiltonian $H_{\textrm{sub}} = -\sum_{G_i \in \mathcal{G}_{\textrm{sub}}} J_i G_i $, and thus consider Hamiltonians containing only some elements from the gauge group $\mathcal{G}_{\textrm{sub}}$.
We emphasize that the symmetry group of the model $H_{\textrm{sub}}$ is the stabilizer group $\mathcal{S}_{\textrm{sub}}$ of the code.
In this subsection we start with three different Hamiltonians associated with the 3D gauge color code, $H^X_{\textrm{GCC}}$, $H^Z_{\textrm{GCC}}$ and $H^Y_{\textrm{GCC}}$, and ungauge their $Z$-type symmetry group generated by the gauge color code $Z$-type stabilizers from $\mathcal{S}_{\textrm{GCC}}$.
The models we consider are trivially equivalent in the absence of symmetries since one we can obtain any of three Hamiltonians from the other two by applying some constant-depth Clifford circuits.
However, in the presence of stabilizer symmetries $H^X_{\textrm{GCC}}$, $H^Z_{\textrm{GCC}}$ and $H^Y_{\textrm{GCC}}$ fall into distinct SPT phases.
In particular, we will see that these three models are mapped to the paramagnet, the 3D toric code and the RBH model.

First, we consider the Hamiltonian
\begin{equation}
H^X_{\textrm{GCC}} = - J_X \sum_{e\in\face 1 {\mathcal{L}}} X(\bnd 1 3 e)
= -J_X \sum_{e\in\face 1 {\mathcal{L}}} X\left( \figbox{1}{eq/eq_star_e}\right).
\end{equation}
which corresponds to the stabilizer Hamiltonian of the 3D stabilizer color code with the $X$-type and $Z$-type stabilizers associated with edges $\face 1 {\mathcal{L}}$ and vertices $\face 0 {\mathcal{L}}$ of the lattice $\mathcal{L}$.
By ungauging the $Z$-type symmetries from $\mathcal{S}_{\textrm{GCC}}$ we find that 
\begin{equation}
H^X_{\textrm{GCC}} \ugmap H_{\textrm{para}} = -J_X \sum_{e\in \face 1 {\mathcal{L}}} X_e = -J_X \sum_{e\in\face 1 {\mathcal{L}}} \figbox{1}{eq/eq_gcc_Xb}.
\end{equation}
which describes a trivial model, the paramagnet.

The other model we consider is defined by the following Hamiltonian
\begin{equation}
H^Z_{\textrm{GCC}} = - J_Z \sum_{e\in\face 1 {\mathcal{L}}} Z(\bnd 1 3 e)
= -J_Z \sum_{e\in\face 1 {\mathcal{L}}} Z\left( \figbox{1}{eq/eq_star_e}\right).
\end{equation}
One can check that $H^Z_{\textrm{GCC}}$ is transformed by the ungauging map $\widetilde\Gamma$ as follows
\begin{equation}
H^Z_{\textrm{GCC}} \ugmap H_{\textrm{TC}} = -J_Z \sum_{e\in \face 1 {\mathcal{L}}} Z(\link 1 e )
= -J_Z \sum_{e\in\face 1 {\mathcal{L}}} Z\left( \figbox{1}{eq/eq_link_e}\right).
\end{equation}
The final $X$-type symmetry group $\mathcal{S}_{\textrm{fin}}$ of the model is specified by Eq.~(\ref{eq_LGT_fin}).
We observe that the final model has local $1$-form symmetries of $X$-type.
Thus, the Hamiltonian $H_{\textrm{TC}}$ describes six copies of the 3D toric code, each of which is supported on a different sublattice $\mathcal{L}^{ab}$.

Lastly, we choose the Hamiltonian
\begin{equation}
H^Y_{\textrm{GCC}} = - J_Y \sum_{e\in\face 1 {\mathcal{L}}} Y(\bnd 1 3 e) = -J_Y \sum_{e\in\face 1 {\mathcal{L}}}
Y\left( \figbox{1}{eq/eq_star_e}\right).
\end{equation}
whose $Y$-type terms are the products of corresponding $X$-type and $Z$-type terms of $H^X_{\textrm{GCC}}$ and $H^Z_{\textrm{GCC}}$.
After ungauging the $Z$-type symmetries we get the following Hamiltonian
\begin{equation}
H^Y_{\textrm{GCC}} \ugmap H_{\textrm{RBH}} = -J_Y \sum_{e\in \face 1 {\mathcal{L}}} X_e Z(\link 1 e )
= -J_Y \sum_{e\in\face 1 {\mathcal{L}}} \figbox{1}{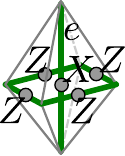}
\end{equation}
The final $X$-type symmetry group $\mathcal{S}_{\textrm{fin}}$ of the model is specified by Eq.~(\ref{eq_LGT_fin}).
We remark that the Hamiltonian $H_{\textrm{RBH}}$ describes three copies of the RBH model, each of which is defined on a pair of sublattices $\mathcal{L}^{ab} \cup \mathcal{L}^{cd}$ (similarly for other two pairs $\mathcal{L}^{ac} \cup \mathcal{L}^{bd}$ and $\mathcal{L}^{ad} \cup \mathcal{L}^{bc}$).

We remark that the RBH model, which was originally studied in the context of measurement-based quantum computation~\cite{Raussendorf:2005aa}, may look similar to the 3D toric code.
However, the RBH model differs in a significant manner due to the presence of symmetries~\cite{Beni15b}.
Let us consider one copy of the RBH model supported on the pair of sublattices $\mathcal{L}^{ab} \cup \mathcal{L}^{cd}$; a similar discussion follows for the other pairs of sublattices $\mathcal{L}^{ac} \cup \mathcal{L}^{bd}$ and $\mathcal{L}^{ad} \cup \mathcal{L}^{bc}$.
The RBH model is an example of a non-trivial SPT order which is protected by local $1$-form symmetries of $X$-type, $X(\bnd 0 1 ^{ab} v)$ and $X(\bnd 0 1 ^{cd} v)$
The triviality of the ground state of the RBH model without any imposed symmetries follows from the fact that there exists a constant-depth disentangling circuit $U_{\textrm{dis}}$ built of the controlled-$Z$ gates acting on qubits on the opposite edges of color $ab$ and $cd$ for every tetrahedron $t$ in the lattice
\begin{equation}
U_{\textrm{dis}} = \prod_{t\in\face 3 {\mathcal{L}}} \figbox{1}{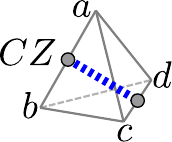}.
\end{equation}
The unitary $U_{\textrm{dis}}$ transforms the ground state of the RBH model it into a trivial product state.
At the same time one can show that the ground state of the RBH is non-trivial in the presence of $1$-form $(\mathbb{Z}_{2})^2$ symmetries.

\subsection{Thermal stability of the 3D gauge color code}

Finally, we discuss the thermal stability of the 3D GCC.
Using to the aforementioned ungauging procedure, we can address this problem by studying thermal properties of the RBH model in the presence or absence of symmetries.
While the RBH model is thermally trivial without any finite temperature phase transition, it had been long suspected that the RBH model exhibits some kind of thermal stability which cannot be directly seen from properties of the thermal Gibbs ensemble $\rho_{\beta}$.
One compelling evidence was the existence of a certain transition temperature $T_{c}>0$ below which $\rho_{\beta}$ serves as a universal resource state for measurement-based quantum computation \cite{Raussendorf:2005aa}. 
This expectation has been confirmed positively in Ref.~\cite{Roberts:2017ab} by showing that the RBH model indeed possesses thermal order in the presence of $1$-form symmetries.
Namely, by restricting the Hilbert space to wavefunctions satisfying $1$-form symmetries, the RBH model undergoes a phase transition at a finite temperature.
We remark that while $1$-form symmetries may not emerge naturally in actual physical systems, they can be effectively imposed by performing active error correction.

Similarly to the RBH model, the 3D gauge color code has been believed to also possess some sort of thermal stability.
In particular, a suggestive evidence of thermal stability is the single-shot error correction property \cite{Bombin:2015aa}, which is typically exhibited by self-correcting quantum memories such as the 4D toric code.
At the same time, the 3D stabilizer color code, which is included the codeword space $\mathcal{H}(\mathcal{S}_{\textrm{GCC}})$ of the gauge color code,
is not topologically ordered at any finite temperature as it is equivalent to multiple copies of the three-dimensional toric code~\cite{Kubica15b}.
Our results offer a potential resolution to this puzzle concerning thermal stability of the gauge color code.
In the absence of symmetries (corresponding to the stabilizer group $\mathcal{S}_{\textrm{GCC}}$), stabilizer Hamiltonians in the 3D GCC are not thermally stable.
Namely, their thermal Gibbs ensemble can be prepared efficiently from some classical ensembles. 
In contrast, in the presence of symmetries the 3D Hamiltonian $H^Y_{\textrm{GCC}}$ is thermally stable due to the thermal stability of the RBH model under the emergent symmetries.

We emphasize that most of previous works on the classification of quantum phases have focused on long-range entanglement at $T=0$. This is partly due to the fact that all the known 2D and 3D topological phases undergo a thermal phase transition at $T_{c}=0$ in the absence of additional symmetries.
Our findings on the other hand open new research avenues for the classification of thermal phases.
Three models related to the gauge color code which we discussed, i.e., 
$H^X_{\textrm{GCC}}$, $H^Z_{\textrm{GCC}}$ and $H^Y_{\textrm{GCC}}$,
should be considered as representatives of distinct thermal SPT phases of matter as their symmetric Gibbs ensembles cannot be transformed into one another via short-depth symmetric circuits.
It is thus tempting to speculate that fault-tolerant properties of the gauge color code stem from an SPT order at finite temperature in the presence of symmetries enforced by active quantum error correction.

We have seen that some stabilizer Hamiltonians\footnote{We remark that there is always some subtlety in choosing stabilizer generators included in a stabilizer Hamiltonian even in the case of a stabilizer code.} associated with the GCC are thermally stable in a sense of their ground state (as well as the thermal Gibbs ensemble) remaining non-trivial up to some non-zero temperature.
As such, the separation of quantum phases associated with the GCC extends to non-zero temperature.
We emphasize that the presence of symmetries is essential in this argument and choosing stabilizer symmetries essentially restricts the allowed states to the codeword space of the GCC.

\section{Fracton symmetry-protected topological phase}
\label{sec_fracSPT}

The framework of ungaguging we discussed in Sec.~\ref{sec_ungauging} applies to arbitrary CSS codes.
A particularly intriguing but at the same time not well-understood class of CSS stabilizer codes is the family of three-dimensional fracton codes with logical operators forming fractal-like objects.
A lot of insight into physics related to fracton models can gained by the ungauging procedure~\cite{Vijay:2016aa, Williamson:2016aa}.
Our ungauging procedure based on the chain complex is not limited to stabilizer codes, and thus provides a systematic way of ungauging symmetries for a potentially much wider family of CSS fracton codes.
In this section, we focus on on a particular CSS stabilizer code, the 3D fractal code~\cite{Beni11b}, which supports logical operators in the shape resembling the Sierpi\'nski triangle.
Inspired by the 3D fractal code, we will introduce the notion of fracton symmetry-protected topological (frac-SPT) phases, which are a generalization of SPT phases protected by fractal-like symmetries.
We use the abbreviation ``frac-SPT'' since fSPT is often used for Floquet SPT phases.
Lastly, we provide an explicit example of a 2D fracton SPT phase and show its non-triviality in the presence of fractal-like symmetries.

\subsection{3D fractal code}

We briefly discuss the 3D fractal code.
Let us consider a three-dimensional $L\times L \times L$ cubic lattice $\mathcal{L}_{\textrm{cub}}$ with periodic boundary conditions.
There are two qubits, labeled by $A$ and $B$, placed at each vertex of the lattice.
We denote by $P_v^A$ and $P_v^B$ a Pauli operator $P\in \{ X,Z\}$ supported on the qubit $A$ or $B$ at the vertex $v \in \face 0 {\mathcal{L}_{\textrm{cub}}}$.
The stabilizer group $\mathcal{S}_{\textrm{frac}}$ of the 3D fractal code is given by
\begin{equation}
\mathcal{S}_{\textrm{frac}} = \left\langle S_v^X, S_v^Z \middle | \forall v \in \face 0 {\mathcal{L}_{\textrm{cub}}}\right\rangle
= \left\langle \figbox{1}{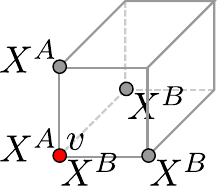}, \figbox{1}{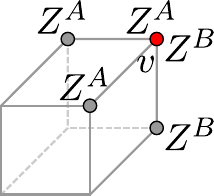} \right\rangle,
\end{equation}
where $S_v^X$ and $S_v^Z$ denote two stabilizers of $X$- and $Z$-type associated with the vertex $v$.
One can check that the logical operators are in the shape of strings and Sierpi\'nski fractals.

We would like to ungauge the initial $Z$-type symmetry group $\sini$, which is generated by the $Z$-type symmetries from $\mathcal{S}_{\textrm{frac}}$ and the logical $\overline Z$ operators.
We remark that the logical $\overline Z$ operators of the 3D fractal code can be represented as $Z^A$-type vertical string-like operators, as well as $Z^B$-type horizontal fractal-like operators supported in every layer $l=1,\ldots, L$.
Thus, we arrive at the following
\begin{equation}
\sini = \left\langle \figbox{1}{eq/eq_fractal_Z}\ , \figbox{1}{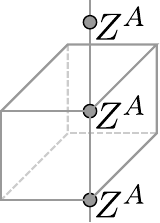}\ ,\ \  Z^B\left(\figbox{.8}{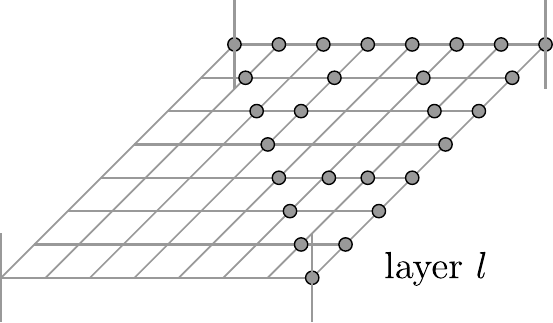}\right) \right\rangle.
\label{eq_some_random_eq}
\end{equation}
We remind the reader that in Eq.~(\ref{eq_some_random_eq}) we use the notation $Z^B(\cdot)$ to represent a $Z^B$-type operator supported on all the qubits depicted in the schematic.
The symmetric Pauli subgroup $\mathcal{P}(\sini)$ is generated by single qubit $Z^A$ and $Z^B$ operators, as well as by $X$-type stabilizer generators from $\mathcal{S}_{\textrm{frac}}$.
By following the ungauging procedure we find that the ungauging map $\widetilde\Gamma$ transforms symmetric operators from $\mathcal{P}(\sini)$ in the following way
\begin{eqnarray}
\figbox{1}{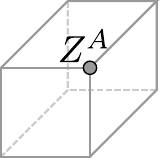} \ugmap \figbox{1}{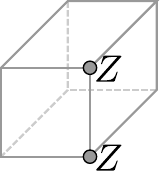},\qquad
\figbox{1}{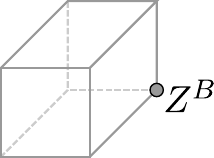} \ugmap \figbox{1}{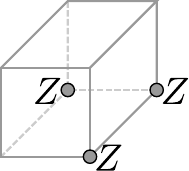},\qquad
S_v^X \ugmap X_v.
\label{eq:op-map}
\end{eqnarray}
Note that the final model has one qubit placed on each and every vertex of the lattice.
After ungauging the stabilizer Hamiltonian $H_{\textrm{frac}}$ associated with the 3D fractal code is mapped to the paramagnetic Hamiltonian
\begin{equation}
H_{\text{frac}} = - \sum_{v\in \face 0 {\mathcal{L}_{cub}}} \figbox{1}{eq/eq_fractal_X}
- \sum_{v\in \face 0 {\mathcal{L}_{cub}}} \figbox{1}{eq/eq_fractal_Z}
\ugmap -\sum_{v\in \face 0 {\mathcal{L}_{cub}}} X_v
\end{equation}
The emergent $X$-type symmetry group $\sfin$ of the model
~\footnote{Strictly speaking, the $X$-type operators depicted in Eq.~(\ref{eq_frac_Xsym}) are not symmetries for any finite $L$ since the $\mathbb{Z}_2$ Sierpi\'{n}ski triangle cannot cover a 2D torus in a consistent manner.
This technicality arises due to the fact that the cellular automaton generating the $\mathbb{Z}_{2}$ Sierpi\'{n}ski triangle is irreversible.
One can avoid this subtlety by considering the 3D fractal code for $m$-dimensional qudits with odd $m>2$.
Another possible resolution is to consider the 3D fractal code on the lattice with boundary; see~\cite{Beni11b} for more details.}
is generated by $X$-type operators, each of which is a product of fractal-like operators supported within every horizontal layer $l=1,\ldots, L$ of the cubic lattice $L\times L \times L$, namely
\begin{equation}
\sfin = \left\langle \prod_{l=1}^L
X\left( \figbox{.8}{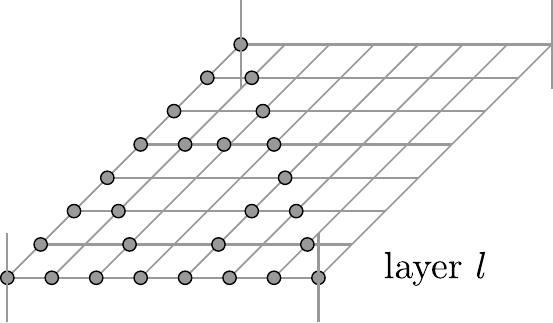} \right)
\right\rangle
\label{eq_frac_Xsym}
\end{equation}

\subsection{Gapped domain wall}

We have seen that fractal symmetries appear naturally as a result of ungauging the 3D fractal code.
This observation motivates us to introduce a new notion of fracton symmetry-protected topological (frac-SPT) phases which would be short-range entangled if no symmetries are imposed and become non-trivial in the present of fractal-like symmetries.
In the reminder of this section, we construct an explicit example of the 2D frac-SPT phase.
We achieve this goal by ungauging a gapped domain wall arising in the aforementioned 3D fractal code.
This subsection discusses a construction of a gapped domain wall from transversal logical operators in the 3D fractal code.

First, let us consider a code dual to the 3D fractal code.
The dual 3D fractal code is also defined on the cubic lattice, with two qubits labeled by $\tilde A$ and $\tilde B$ placed on every vertex.
The stabilizer group $\tilde{\mathcal{S}}_{frac}$ of the dual 3D fractal code is given by
\begin{equation}
\tilde{\mathcal{S}}_{\text{frac}} = \left\langle \tilde{S}_v^X, \tilde{S}_v^Z \middle | \forall v \in \face 0 {\mathcal{L}_{\text{cub}}}\right\rangle
= \left\langle 
\figbox{1.0}{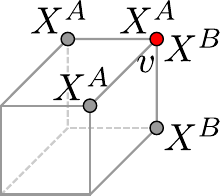}, \figbox{1.0}{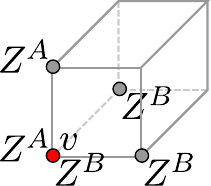} \right\rangle.
\end{equation}
In other words, the stabilizers of the dual code are obtained from the stabilizers of the 3D fractal code by applying the transversal Hadamard, which swaps Pauli $X$ and $Z$ operators.

Let us consider the 3D fractal code and the dual code combined.
We define the transversal controlled-$Z$ gate $\overline{CZ}$ to be a product of the controlled-$Z$ gates implemented between every pair of corresponding qubits, one from the fractal code and the other from the dual code.
We will show that $\overline{CZ}$ is a logical operation on both codes.
To see that, recall that the $CZ$ gate commutes with Pauli $Z$ operators and transforms Pauli $X$ operators in the following way
\begin{equation}
\text{C}Z (X\otimes I) \text{C}Z^\dag = X\otimes Z, \qquad \text{C}Z (I\otimes X) \text{C}Z^\dag = Z\otimes X.
\end{equation}
Thus, the transversal $\text{C}Z$ gate preserves the $Z$-type stabilizers of both codes and transforms the $X$-type stabilizer generators as follows
\begin{align}
\overline{\text{C}Z} (S_{v}^{X} \otimes I ) \overline{\text{C}Z}^\dag = S_{v}^{X} \otimes\tilde{S}_{v}^{Z}, \qquad 
\overline{\text{C}Z} (I \otimes \tilde{S}_{v}^{X} ) \overline{\text{C}Z}^\dag =S _{v}^{Z} \otimes \tilde{S}_{v}^{X},
\label{eq:CZ-fracton}
\end{align}
i.e., the $X$-type stabilizers are decorated by $Z$-type stabilizers from the dual code, and vice versa.
Therefore, the action of $\overline{\text{C}Z}$ preserves the codeword space, and thus is a logical operator.
In fact, one can show that $\overline{\text{C}Z}$ is a non-trivial logical operator.

\begin{figure}[h!]
\centering
\includegraphics[width=.3\columnwidth]{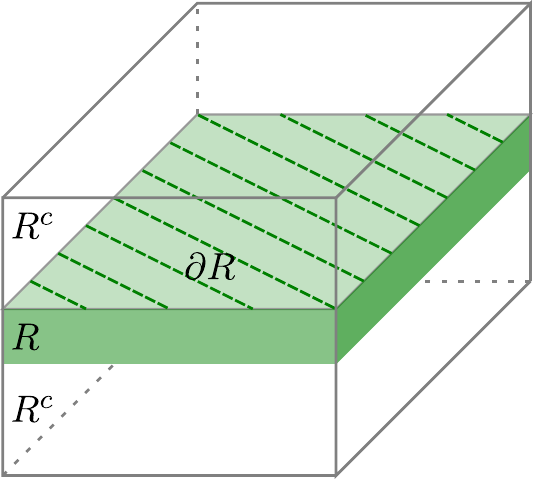}
\caption{
A construction of a gapped domain wall in a model consisting of two copies of the 3D fractal code, $H^{\textrm{tot}}_{\textrm{frac}} = H_{\textrm{frac}} + \tilde{H}_{\textrm{frac}}$.
By applying the transversal $\overline{CZ}|_R$ operator within the region $R$ (a horizonal slab shaded in green) we can form a gapped domain wall on the surface $\partial R$ (hatched) separating the region $R$ and its complement $R^c$.
}
\label{fig_domain_wall} 
\end{figure}

Next, we construct a gapped domain wall by using a transversal logical gate $\overline{\text{C}Z}$.
Consider a cubic lattice with four qubits labeled by $A$, $B$, $\tilde A$ and $\tilde B$ placed at each vertex.
The qubits labeled by $A$, $B$ are associated with the 3D fractal code, whereas the qubits $\tilde A$, $\tilde B$ are identified with the dual code.
We consider the following 3D Hamiltonian 
\begin{eqnarray}
H^{\text{tot}}_{\text{frac}} = H_{\text{frac}} + \tilde{H}_{\text{frac}} 
\end{eqnarray}
where $H_{\text{frac}}$ and $\tilde H_{\text{frac}}$ are the stabilizer Hamiltonians for the 3D fractal code and the dual code.
We emphasize that we include the $Z$-stabilizers in $H_{\text{frac}}$ and $\tilde H_{\text{frac}}$.
Note that $H_{\text{frac}}$ and $\tilde{H}_{\text{frac}}$ are not interacting.
We transform the 3D Hamiltonian $H^{\text{tot}}_{\text{frac}}$ by implementing the transversal control-$Z$ operator only within some region $R$ of the lattice, as depicted in Fig.~\ref{fig_domain_wall}.
We emphasize that the region $R$ is chosen in such a way that its surface $\partial R$ is a horizontal 2D sheet.
Then, we obtain
\begin{equation}
(\overline{\text{C}Z}|_{R}) (H^{\textrm{tot}}_{\text{frac}}) (\overline{\text{C}Z}|_{R})^\dag = H_{R} + H_{\partial{R}} + H_{R^c},
\label{eq_htot_cz}
\end{equation}
where $\overline{\text{C}Z}|_{R}$ denotes the restriction of $\overline{\text{C}Z}$ onto $R$.
We have grouped terms of the resulting Hamiltonian in Eq.~(\ref{eq_htot_cz}) into three parts $H_{R}, H_{\partial{R}}, H_{R^c}$.
The Hamiltonians $H_{R}$ and $H_{R^c}$ contain terms supported inside either the region $R$ or its complement $R^c$, respectively, whereas $H_{\partial{R}}$ includes all the terms which are supported jointly over $R$ and $R^c$.
Since $\overline{\text{C}Z}|_{R}$ is constant-depth geometrically local operator, the terms in $H_{\partial{R}}$ are localized around the surface $\partial R$ separating $R$ from its complement $R^c$.

A crucial observation is that $\overline{\text{C}Z}|_{R}$ has an effect of creating a transparent gapped domain wall
\footnote{By a transparent domain wall, we understand a 2D surface separating two regions, such that no excitation can be created or annihilated there.  However, excitations of the model can be transformed in some non-trivial way upon crossing the domain wall.}
described by the Hamiltonian $H_{\partial R}$ supported on the surface $\partial R$.
Let us examine how the terms in the Hamiltonian $H^{\textrm{tot}}_{\textrm{frac}}$ are transformed by $\overline{\text{C}Z}|_{R}$.
Inside the region $R$, the $X$-type stabilizers $S_v^X$ and $\tilde{S}_v^X$ are transformed respectively into $S_{v}^{X} \otimes \tilde{S}_{v}^{Z} $ and $S_{v}^{Z} \otimes \tilde{S}_{v}^{X}$, as in Eq.~\eqref{eq:CZ-fracton}.
However, $X$-type stabilizers in $H_{\partial R}$, localized around the surface $\partial R$ are transformed in some non-trivial manner since $\overline{CZ}|_R$ gate acts only on a part of their support.
Finally, terms in $H_{R^c}$ remain unchanged.
Since terms in $H_{\partial{R}}$ commute with all the stabilizers supported on $R$, one can replace $S_{v}^{X} \otimes \tilde{S}_{v}^{Z} $ and $S_{v}^{Z} \otimes \tilde{S}_{v}^{(X)}$ in $H_{R}$ by $S_{v}^{(X)} \otimes I $ and $I \otimes \tilde{S}_{v}^{(X)}$ without changing the ground space of the overall Hamiltonian.
Thus, we arrive at the following Hamiltonian
\begin{align}
H^{\text{DW}}_{\text{frac}} = H'_{R} + H_{\partial R} + H_{R^{c}}.
\end{align}
where $H'_{R}$ and $H_{R^{c}}$ consist of terms identical to those in the original Hamiltonian $H^{\textrm{tot}}_{\text{frac}}$ and $H'_{R}$ is obtained by the above replacement of stabilizer generators.
Note that the 3D fractal code and the dual code are coupled only at $\partial R$ in $H_{\text{DW}}$.
Therefore, the Hamiltonian $H_{\partial R}$ can be viewed as describing a gapped domain wall on the surface $\partial R$ in between two regions $R$ and $R^c$.

The gapped domain wall we constructed exchanges $X$-type and $Z$-type fracton excitations in a non-trivial manner since $\overline{\text{C}Z}$ is a non-trivial logical gate.
Note that the $X$-type and $Z$-type fracton excitations of the model have non-trivial braiding statistics which cannot be altered via any local unitary acting within the neighborhood of $\partial R$. 
This implies that the domain wall cannot be created by applying a geometrically local constant-depth unitary around  $\partial R$; see~\cite{Beni15} for a detailed discussion.

\subsection{2D fracton symmetry-protected topological phase}

Now we are going to present an explicit example of the 2D frac-SPT phase, which is protected by the $\mathbb{Z}_2\times \mathbb{Z}_2$ fractal-like symmetry in the shape of the Sierpi\'nski triangle. 
In the previous subsection, we have constructed the Hamiltonian $H^{\text{DW}}_{\text{frac}}$ with a gapped domain wall by appropriately modifying terms near the surface $\partial R$ in the combined Hamiltonian $H_{\textrm{frac}}^{\textrm{tot}} = H_{\text{frac}} + \tilde{H}_{\text{frac}}$ of the 3D fractal code and the dual code.
Now we will see that the 2D fracton SPT Hamiltonian $H_{\text{frac-SPT}}$ can be obtained by ungauging the Hamiltonian $H^{\text{DW}}_{\text{frac}}$.
Namely, let the initial Hilbert space $\mathcal{H}_{\textrm{ini}}$ be identified with the qubits of the model $H^{\text{DW}}_{\text{frac}}$ with the domain wall, i.e., $\mathcal{H}_{\textrm{ini}}$ is associated with four qubits (labeled $A$, $B$, $\tilde A$, $\tilde B$) at every vertex of the lattice $\mathcal{L}_{\textrm{cub}}$.
We can consider the ungauging procedure which transforms between $\mathcal{H}_{\textrm{ini}}$ and the final Hilbert space $\mathcal{H}_{\textrm{fin}}$ associated with two qubits (labeled by $I$, $II$) per vertex. The labels $I$ and $II$ indicate which copy of the 3D fractal code each qubit of the 2D frac-SPT model is associated with. 
Namely, we want to ungauge the $Z$-symmetries of the combined the 3D fractal code and the dual code.
The ungauging map $\widetilde\Gamma$ maps the Hamiltonian $H^{\text{DW}}_{\text{frac}}$ with a gapped domain wall as follows
\begin{equation}
H^{\text{DW}}_{\text{frac}} \ugmap H_{\textrm{para},R} + H_{\textrm{para},R^c} + H_{\text{frac-SPT},\partial R},
\label{eq_DW_SPT}
\end{equation}
where $H_{\text{para}, R}$ and $H_{\text{para}, R^c}$ are paramagnetic Hamiltonians on all qubits $I$, $II$ within the region $R$ and its complement $R^c$, respectively.

We claim that the last term $H_{\text{frac-SPT},\partial R}$ in Eq.~(\ref{eq_DW_SPT}), supported on the horizontal surface $\partial R$, describes the 2D frac-SPT Hamiltonian. 
Let us explicitly write down $H_{\text{frac-SPT},\partial R}$ on the square lattice $\mathcal{L}_{sq}$ of size $L\times L$ with periodic boundary conditions.
There are two qubits, labeled $I$ and $II$, placed on each vertex of the lattice.
The Hamiltonian describing a non-trivial 2D frac-SPT phase is given by
\begin{equation}
H_{\text{frac-SPT}} = - \sum_{f \in \face 2 {\mathcal{L}_{sq}}} \figbox{1.0}{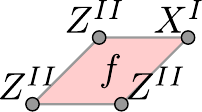}
- \sum_{f \in \face 2 {\mathcal{L}_{sq}}}  \figbox{1.0}{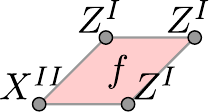},
\label{eq_H_fracSPT}
\end{equation}
where the sum is over all faces $f\in \face 2 {\mathcal{L}_{sq}}$ of the square lattice.
One may verify that all the terms in the Hamiltonian commute with one another.
Moreover, they also commute with the elements of the following fractal symmetry group
\begin{eqnarray}
\label{eq_fSPT_symmetry}
\mathcal{S}^X_{\textrm{frac}} = \left\langle 
X^I\left(\figbox{.8}{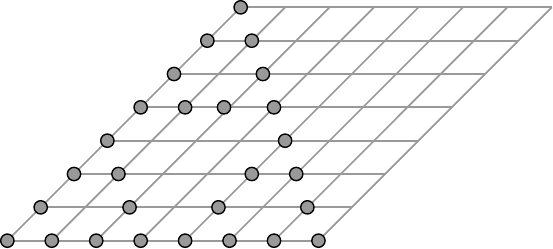}\right),
X^{II}\left(\figbox{.8}{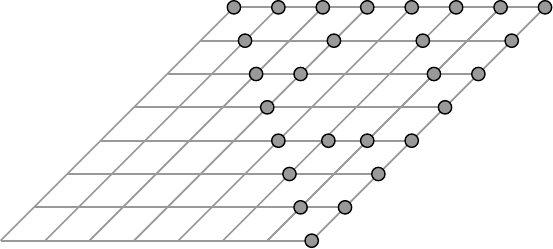}\right)\right\rangle.
\end{eqnarray}
We emphasize that $\mathcal{S}^X_{\textrm{frac}}$ is generated by the $X$-type operators depicted in Eq.~(\ref{eq_fSPT_symmetry}) as well as their translations.
The appearance of the 2D Sierpi\'nski fractals symmetries in $H_{\text{frac-SPT},\partial R}$ can be understood as follows. Observe that the right-hand side of Eq.~(\ref{eq_DW_SPT}) commutes with 3D fractal symmetry operators as depicted in Eq.~(\ref{eq_frac_Xsym}). Note that the restriction of the full 3D symmetry operators to the surface $\partial R$ are the 2D Sierpi\'nski fractals. As such, $H_{\text{frac-SPT},\partial R}$ must commute with the 2D Sierpi\'nski fractal operators. 

We remark that the Hamiltonian $H_{\text{frac-SPT}}$ is trivial in the absence of fractal symmetries from $\mathcal{S}^X_{\text{frac-SPT}}$.
Namely, $H_{\text{frac-SPT}}$ can be transformed into a paramagnetic Hamiltonian by applying the following constant-depth local unitary $U_{dis}$ built of the controlled-$Z$ gates
\begin{equation}
U_{dis} = \prod_{f \in \face 2 {\mathcal{L}_{sq}}} CZ\left(\figbox{1.0}{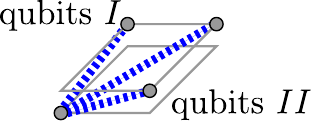}\right)
\end{equation}
where blue dashed lines depict local $CZ$ gates between qubits $I$ and $II$.
It is worth noting that the Hamiltonian $H_{\text{frac-SPT}}$ supports a cluster state as a ground state.

We point out that $H_{\text{frac-SPT}}$ can be defined for a lattice with open boundary conditions in the $\hat y$ direction.
In that case, there will be dangling degenerate boundary modes  protected by the bulk fractal symmetries, which is an evidence of non-triviality of the SPT phase of $H_{\text{frac-SPT}}$.

Finally, let us prove the non-triviality of the Hamiltonian $H_{\text{frac-SPT}}$.
Suppose that $H_{\text{frac-SPT}}$ is trivial, i.e., there exists a constant-depth local disentangling unitary $U_{\textrm{dis}}$ composed of gates commuting with the symmetries which transforms the ground state of  $H_{\text{frac-SPT}}$ into a product state.
Note that the unitary $\Gamma (U_{\textrm{dis}})$, which is obtained by applying the gauge map $\Gamma$ to $U_{\textrm{dis}}$, is also constant-depth and local.
At the same time, by applying $\Gamma (U_{\textrm{dis}})$ we can create the domain wall on the surface $\partial R$ in the Hamiltonian $H^{\text{DW}}_{\text{frac}}$.
This is a contradiction with a fact that the gapped domain wall Hamiltonian $H^{\text{DW}}_{\text{frac}}$ is non-trivial.
Thus, we conclude that there is no constant-depth symmetric disentangling unitary for $H_{\text{frac-SPT}}$.

We remark that it is possible to consider an arbitrary subregion $R$ whose surface $\partial R$ is not necessarily horizontal.
Similarly, the fracton SPT Hamiltonian would also appear by ungauging the domain wall on $\partial R$.
The symmetry operators could be found in that case by restricting the full symmetries to $\partial R$.

\subsection{A general prescription for $D$-dimensional SPT phases}
\label{sec_generalized_recipe}

It is worth emphasizing that the presented construction of the 2D fracton SPT Hamiltonian in the previous subsection can be applied to arbitrary CSS stabilizer codes.
For the convenience of readers we summarize this procedure. 

\begin{enumerate}
\item Given a $(D+1)$-dimensional CSS stabilizer code $\mathcal{C}$ with geometrically-local generators, construct its dual code $\tilde{\mathcal{C}}$ by applying the transversal Hadamard gate, which swaps all $X$-type and $Z$-type stabilizers.
\item Construct the tensor-product code $\mathcal{C} \otimes \tilde{\mathcal{C}}$ which combines of the initial and the dual codes.
\item Observe that the transversal gate $\overline{\text{C}Z}$ between pairs of corresponding qubits from $\mathcal{C}$ and $\tilde{\mathcal{C}}$ is a non-trivial logical gate for the code $\mathcal{C} \otimes \tilde{\mathcal{C}}$.
\item Starting from the stabilizer Hamiltonian for $\mathcal{C} \otimes \tilde{\mathcal{C}}$ construct the Hamiltonian $H_{\text{DW}}$ with a gapped domain wall by applying the transversal gate $\overline{\text{C}Z}|_R$ only within some region $R$. 
\item Follow the procedure of ungauging to ungauge the $Z$-type symmetries of the code $\mathcal{C} \otimes \tilde{\mathcal{C}}$.
\item Apply the ungauging map $\widetilde\Gamma$ to $H_{\text{DW}}$ to obtain a $D$-dimensional SPT Hamiltonian $H_{\textrm{SPT}}$ supported on the surface $\partial R$ of the region $R$.
\item Symmetries of the SPT Hamiltonian $H_{\textrm{SPT}}$ can be found by restricting full symmetry operators in the ungauged system to $\partial R$.
\end{enumerate}

We remark that we can apply this method to the 2D toric code.
Then, we obtain the 1D cluster state, which is SPT ordered under $0$-form $\mathbb{Z}_{2}\times \mathbb{Z}_{2}$ symmetries. 
In principle, we could also apply the above prescription to an arbitrary CSS subsystem code, which would generate SPT phases protected by subsystem symmetries, similar to the ones proposed in~\cite{You18}.
However, in that case the resulting domain wall is not guaranteed to be gapped.
It is likely that such approaches will provide a number of interesting quantum phases of matter and gapless systems protected by exotic symmetries.

\section{Discussion}

In this paper we have discussed the procedures of gauging and ungauging viewed from the perspective of quantum error correction and their implications for quantum phases of matter.
Gauging and ungauging provide valuable insights into a question of whether different stabilizer Hamiltonians associated with the subsystem code represent different phases of matter in the presence of stabilizer symmetries.
Our results suggest that phase distinction at finite temperature is essential for fault-tolerant code switching between subspaces of different stabilizer Hamiltonians when syndrome measurements are faulty~\cite{Paetznick13, Kubica15, Bombin:2015aa,Bombin:2016aa, Bombin15}.

Our work has focused on gauging and ungauging Pauli CSS symmetries.
An exciting generalization would be to consider procedures of gauging and ungauging applicable to non-CSS subsystem codes.
We remark that the two-dimensional Kitaev honeycomb model~\cite{Kitaev06b} can be interpreted as a non-CSS subsystem code generated by two-body $X$-, $Y$- and $Z$-type Pauli operators (depending on the orientation of the edge).
Its codeword subspace protected by non-CSS stabilizer symmetries can be obtained from a symmetric subspace of the fermionic Hilbert space by gauging the global fermionic parity symmetry.
Note that this idea was originally used to solve the model.
We believe that gauging and ungauging associated with non-CSS subsystem codes may provide useful insights into fermionic SPT phases.

Our analysis of CSS subsystem codes naturally led to a procedure of partial gauging, i.e., given the full symmetry group $G$ of the system gauge only some of its symmetries from the subgroup $H$ of $G$.
The idea of partial gauging has recently been used to explain certain paradoxical construction of a gapped boundary in a two-dimensional SPT phase where the $\mathbb{Z}_{2}$ subgroup of the full $\mathbb{Z}_{4}$ symmetry was gauged \cite{WWW17}.
{We speculate that} such systems may {be interpreted as subsystem quantum error-correcting codes.}
Another concept relevant to partial gauging is symmetry-enriched topological (SET) phases of matter which describe topological phases protected by additional symmetries \cite{Chen:2017aa, Mesaros:2013aa}. 
In that setting, the preserved symmetries after partial gauging can be interpreted as those additional symmetries in the gauged system.

We have seen that using transversal logical gates in $(D+1)$-dimensional CSS stabilizer codes with geometrically-local generators one can construct $D$-dimensional gapped domain walls.
After ungauging these domain walls can provide explicit examples of non-trivial fracton SPT phases protected by fractal-like symmetries.
We emphasize that it is also possible to reason in the other direction, i.e., by using the theory of $D$-dimensional bosonic $0$-form SPT phases one can find a family of $(D+1)$-dimensional topological codes with transversal logical gates~\cite{Beni15c}.
Interestingly, by explotining similar ideas the classification of Floquet SPT phases has beed developed~\cite{Else:2016aa}.
Thus, a systematic exploration of transversal logical gates in subsystem codes should be intimately related to the (dynamical) classification of SPT phases protected by subsystem symmetries.

We finally remark that all known 2D and 3D models of topological quantum phases of matter have trivial Gibbs ensembles at finite temperature from the viewpoit of thermal stability and quantum circuit complexity~\cite{Beni11, Hastings11, Siva16}.
However, imposing additional symmetries may lead to some non-trivial Gibbs ensembles as already found in the RBH model and the gauge color code.
The quantum circuit complexity characterization of exotic critical points (such as the critical point for the gauge color code) in the presence of stabilizer symmetries is an interesting problem.
The Gibbs ensemble at quantum criticality would be trivial under conventional global symmetries as it is short-range correlated and admits a finite-depth faithful MERA representation~\cite{Evenbly:2015aa}.
The same question in the presence of generalized global symmetries remains open.

\subsection*{Acknowledgements}

We thank Stephen Bartlett, H\'ector Bomb\'in, Fernando Pastawski, John Preskill and Sam Roberts for inspiring discussions. 
In particular, we are grateful to H\'ector Bomb\'in for emphasizing a connection between the gauge color code and $\mathbb{Z}_2$ lattice gauge theory, and to Sam Roberts and Stephen Bartlett for sharing the draft of their work on symmetry-protected self-correcting quantum memories \cite{Sam_Stephen}.
AK acknowledges funding provided by the Simons Foundation through the ``It from Qubit'' Collaboration.
Research at Perimeter Institute is supported by the Government of Canada through Industry Canada and by the Province of Ontario through the Ministry of Research and Innovation.

\appendix

\section{Equivalence between the stabilizer color and toric codes revisited}
\label{sec_equiv}

The stabilizer color code in $D$ dimensions has been shown to be unitary equivalent to multiple copies of the $D$-dimensional toric code \cite{Kubica15b}. 
In particular, there exists a geometrically-local unitary circuit $U$ of constant depth which maps stabilizer generators of the color code to stabilizer generators of the toric code.
Now, we revisit this unitary equivalence and reinterpret the mapping between the color and toric codes as a partial ungauging of some symmetries of the color code.

\begin{figure}[h!]
\centering
\includegraphics{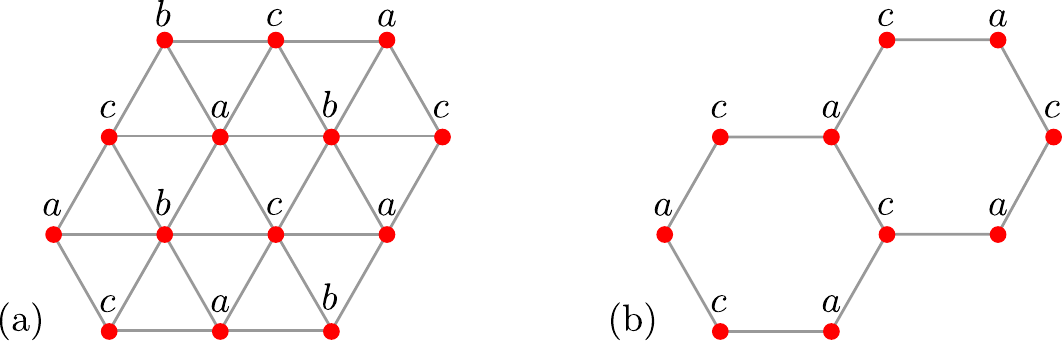}
\caption{
(a) An example of a two-dimensional lattice $\mathcal{L}$ built of triangles. 
We can assign three labels $a$, $b$ and $c$ to the vertices (red dots) in such a way that any two vertices incident to the same edge have different labels.
Note that each face has exactly one vertex with each label.
(b) A sublattice $\mathcal{L}^{ac}$ is obtained from $\mathcal{L}$ by keeping only the vertices of color $a$ or $c$ and the edges between them.
}
\label{fig_2dcc} 
\end{figure}

For simplicity of the discussion, let us consider the 2D stabilizer color code.
The 2D stabilizer color code is defined by placing qubits of the triangular faces of a two-dimensional lattice $\mathcal{L}$ without boundary, whose vertices are colored in three colors $a$, $b$ and $c$; see Fig.~\ref{fig_2dcc}(a).
The $X$- and $Z$-type stabilizer generators of the stabilizer group $\mathcal{S}_{\textrm{CC}}$ are identified with vertices of the lattice, namely
\begin{equation}
\mathcal{S}_{\textrm{CC}} = \left\langle X(\bnd 0 2 v), Z(\bnd 0 2 v) | \forall v\in \face 0 {\mathcal{L}} \right\rangle
= \left\langle X\left(\figbox{.8}{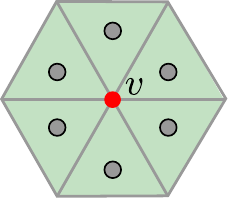}\right), Z\left(\figbox{.8}{eq/eq_star2v}\right) \right\rangle
\end{equation}

We choose to ungauge the $Z$-type symmetry group $\sini$, which is generated only by some $Z$-type stabilizers of the color code.
Namely, we define $\sini$ to be generated by $Z$-type stabilizers associated with vertices of color $c$, i.e.,
$\sini =  \left\langle Z(\bnd 0 2 v) | \forall v\in \face 0 {\mathcal{L}} \wedge \col{v} = c \right\rangle$.
We follow the ungauging procedure described in Sec.~\ref{sec_definition}.
First, we find that the symmetric Pauli subgroup $\mathcal{P}(\sini)$ is generated by single-qubit $Z$ operators, as well as two-qubit $X$ operators $X(\bnd 1 2 e)$ associated with edges $e\in\face 1 {\mathcal{L}}$ of color $ac$ or $bc$, namely
\begin{equation}
\mathcal{P}(\sini) = \langle Z_f, X(\bnd 1 2 e) | \forall f\in \face 2 {\mathcal{L}}, e \in\face 1 {\mathcal{L}} \wedge \col e \in \{ ab, ac \} \rangle.
\end{equation}
The final Hilbert space $\mathcal{H}_{\textrm{fin}}$ is thus identified with the qubits placed on the edges of color $ac$ and $bc$.
The ungauging map $\widetilde\Gamma$ transforms the symmetric Pauli operators from $\mathcal{P}(\sini)$ according to the following prescription
\begin{eqnarray}
\label{eq_cc_gauge1}
\forall f\in \face 2 {\mathcal{L}}: Z_f = \figbox{1.}{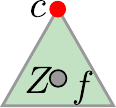} &\ugmap&
\prod_{\substack{e\in \bnd 0 1 v\\ \col e \in \{ac, bc \}}} Z_e = \figbox{1.}{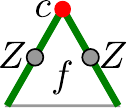},\\
\label{eq_cc_gauge2}
\forall e\in \face 2 {\mathcal{L}} \wedge \col e \in \{ ac, bc\} : X(\bnd 1 2 e) = \figbox{1.}{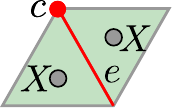} &\ugmap&
 X_e = \figbox{1.}{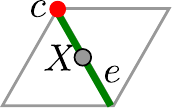}
\label{eq_cc_2dmapping}
\end{eqnarray}

We can verify that the $Z$-type symmetries from $\sini$ are mapped by the ungauging map $\widetilde\Gamma$ to the identity operator.
At the same time, any $Z$-type stabilizer generator of the color code associated with a vertices of color $a$ or $b$ is mapped by $\widetilde\Gamma$ as follows
\begin{equation}
\forall v\in \face 0 {\mathcal{L}} \wedge \col v \in \{b,c\} : Z\left(\figbox{.8}{eq/eq_star2v}\right) \ugmap
\label{eq_Zpres}
Z(\link 1 v) = Z\left(\figbox{.8}{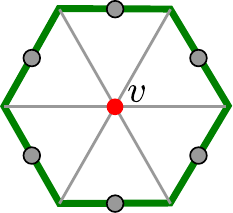}\right).
\end{equation}
We call the $Z$-type operators in the right-hand side of Eq.~(\ref{eq_Zpres}) the preserved $Z$-type symmetries and denote the group they generate by $\mathcal{Z}_{\textrm{pre}}^Z$.
Note that for each vertex $v$ of color $c$ there is a relation between the symmetric $X$-type operators from $\mathcal{P}(\sini)$, i.e., 
$\prod_{e \in \bnd 0 1 v} X(\bnd 1 2 e) = I$.
Thus, the emergent $X$-type symmetry group is generated by $X$-type vertex operators associated with vertices of color $c$, namely
\begin{equation}
\sfin = \langle X(\bnd 0 1 v) | \forall v\in \face 0 {\mathcal{L}} \rangle
= \left\langle X\left(\figbox{.8}{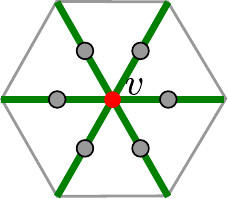}\right) \right\rangle
\end{equation}
We conclude that the final symmetry group
$\mathcal{S}_{\textrm{fin}} = \langle \sfin, \mathcal{S}^Z_{\textrm{pre}}\rangle$ is generated by the $X$-type symmetries from $\sfin$, as well as by the $Z$-type preserved symmetries from $\mathcal{S}^Z_{\textrm{pre}}$.

We can verify the Hamiltonian of the 2D color code 
\begin{equation}
H_{\textrm{CC}} = - \sum_{v\in \face 0 {\mathcal{L}}} X\left(\figbox{.8}{eq/eq_star2v}\right)
\end{equation}
transforms under the ungauging map $\widetilde\Gamma$ as follows
\begin{equation}
H_{\textrm{CC}}  \ugmap H_{\textrm{TC}} 
= - \sum_{\substack{v\in \face 0 {\mathcal{L}}\\ \col v \in \{ a,c \}}}  X\left(\figbox{.8}{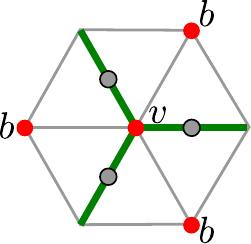}\right)
- \sum_{\substack{v\in \face 0 {\mathcal{L}}\\ \col v \in \{ b,c \}}} X\left(\figbox{.8}{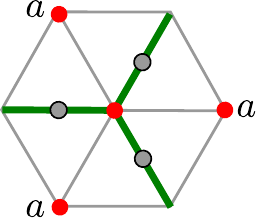}\right)
\label{eq_cc_ungauged}
\end{equation}
and the symmetry group of the ungauged model is $\mathcal{S}_{\textrm{fin}}$.
We identified that the Hamiltonian $H_{\textrm{TC}}$ in Eq.~(\ref{eq_cc_ungauged}) describes two decoupled copies of the 2D toric code.
Note that the copies of the 2D toric code are defined on two sublattices $\mathcal{L}^{ac}$ and $\mathcal{L}^{bc}$.
We remind the reader that the sublattice $\mathcal{L}^{ac}$ (or $\mathcal{L}^{bc}$) is obtained from $\mathcal{L}$ by keeping only the vertices of color $a$ or $c$ (respectively $b$ and $c$) and the edges between them; see Fig.~\ref{fig_2dcc}(b).
We observe that the ungauging map $\widetilde\Gamma$ describes the same transformation between the 2D color code and two copies of the 2D toric code as the one implemented by a local unitary transformation $U$ from Ref.~\cite{Kubica15b}.
In other words, the unitary $U$ transforms the stabilizer generators of the 2D color code into the stabilizer generators of two decoupled copies of the 2D toric code in a way as consistent with Eqs.~(\ref{eq_cc_gauge1})~and~(\ref{eq_cc_gauge2}).

We finally remark that a similar reasoning holds in $D\geq 3$ dimensions.
Let us consider the $D$-dimensional color code of type $k$, where $k \in \{1,2, \ldots, D-1 \}$.
The integer $k$ determines the color code stabilizer group $\mathcal{S}_{\textrm{CC}}$, since the $X$- and $Z$-type stabilizer generators by definition are identified with $(k-1)$- and $(D-k-1)$-simplices in the following way
\begin{equation}
\mathcal{S}_{\textrm{CC}} = \langle X(\bnd {k-1} D \delta), Z(\bnd {D-k-1} D  \sigma) | \delta\in\face{k-1}{\mathcal{L}}, \sigma\in\face{D-k-1}{\mathcal{L}} \rangle
\end{equation}
We can consider ungauging the following $Z$-type symmetry group
\begin{equation}
\sini = \langle Z(\bnd {D-k-1} D  \sigma) | \sigma\in\face{d-k-1}{\mathcal{L}}  \wedge c^*\in \col \sigma\rangle,
\end{equation}
which is generated by $Z$-type stabilizers identified with $(D-k-1)$-simplices containing some chosen color $c^*$.
Then, one can show that the ungauging map $\widetilde\Gamma$ would map between the $D$-dimensional color code and $D \choose k$ copies of the $D$-dimensional toric code, each of which is supported on a different sublattice.
This is in agreement with the unitary equivalence explained in Ref.~\cite{Kubica15b}.

\section{Topological order in the (Majorana) gauge color code}\label{sec:majorana}

We have seen that some of stabilizer Hamiltonians associated with the 3D GCC can be mapped to (multiple copies of) the RBH model. A prominent property of the RBH model is the appearance of surface topological order which is thermally stable when protected by $1$-form symmetries~\cite{Raussendorf:2005aa, Roberts:2017ab}. It is thus natural to ask whether surface topological order also appear in stabilizer Hamiltonians in the 3D GCC too (before ungauging). In this appendix, we discuss two particular classes of stabilizer Hamiltonians which are contained in the three-dimensional gauge color code. We find that one of the models indeed supports the 2D stabilizer color code on its surface whereas the other model possesses intrinsic 3D topological order which is distinct from the 3D stabilizer color code. These findings not only provide another evidence of thermal stability of the 3D GCC without performing ungauging, but also suggest that the symmetric subspace of the 3D GCC can contain a rich variety of interesting topological phases.

A particularly intriguing feature, common in both of the aforementioned models, is that the Hamiltonians are invariant under exchanges of Pauli-$X$ and Pauli-$Z$ operators. This extra symmetry enable us to obtain interesting toy models of topological order made of majorana fermions. This is essentially due to the fact that the CSS self-dual stabilizer Hamiltonian can be automatically converted into a majorana fermion Hamiltonian.

\subsection{Surface topological order}

\begin{figure}[htb!]
\centering
\includegraphics[width=0.35\linewidth]{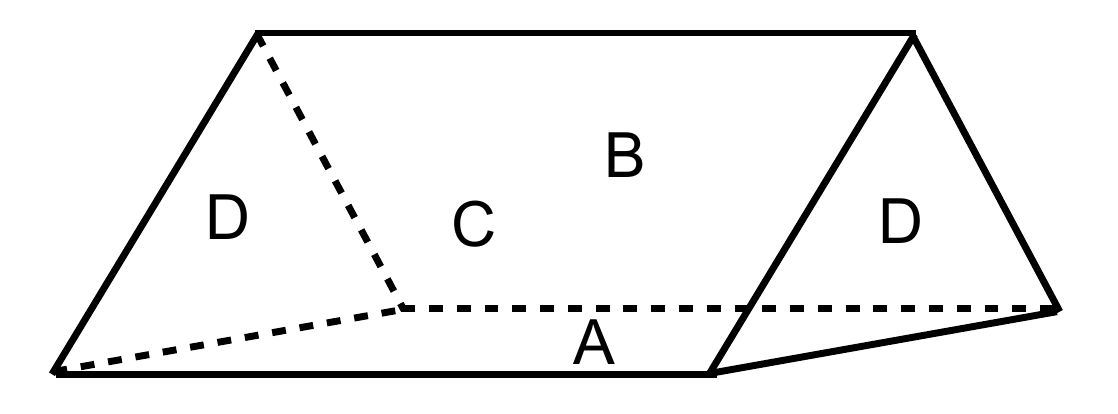}
\caption{Boundary conditions in the gauge color code.
} 
\label{fig_color_boundary}
\end{figure}

The first model we consider has trivial bulk, but supports non-trivial topological order, namely the two two-dimensional color code on the boundaries. Consider the four-colorable lattice in three dimensions. Boundaries are chosen as shown in Fig~\ref{fig_color_boundary} where two opposite boundaries have color label $D$ while three triangular boundaries have color label $A,B,C$ respectively. Let us treat all the stabilizer generators $S_{v}^{X}, S_{v}^{Z}$ as symmetries of the system. Consider the following bulk stabilizer Hamiltonian:
\begin{align}
H_{\text{bulk}} = - \sum_{f_{AD},f_{BD},f_{CD}} ( S^{X}_{f_{AD}} + S^{X}_{f_{BD}} + S^{X}_{f_{CD}} + S^{Z}_{f_{AD}} + S^{Z}_{f_{BD}} + S^{Z}_{f_{CD}}) \label{eq:bulk}
\end{align}
where $f_{AD}, f_{BD}, f_{CD}$ represent $2$-cells of color $AD,BD,CD$. Observe that the above bulk Hamiltonian does not contain any term which couple different $3$-cells of color $D$. This implies that the bulk wavefunction is short-range entangled as different $3$-cells. Yet qubits on the boundaries of color $D$ are decoupled from the bulk, and satisfy stabilizer conditions of the two-dimensional color code localized on the boundaries. Namely, two copies of the two-dimensional color code live on the opposite boundaries. Due to the stabilizer symmetries on the system, the symmetric wavefunction must correspond to maximally entangled ground states of two copies of the two-dimensional color code. The presence of surface topological order implies that stabilizer symmetries in the GCC may act in a way similar to the ordinary $1$-form symmetries which prohibits point-like excitations.

A self-dual CSS stabilizer code can be converted into a Majorana fermion code by replacing Pauli operators with Majorana operators. Let us construct a Majorana version of Eq.~\eqref{eq:bulk}:
\begin{align}
\hat{H}_{\text{bulk}} = - \sum_{f_{AD},f_{BD},f_{CD}} ( K_{f_{AD}} + K_{f_{BD}} + K_{f_{CD}} ) \label{eq:bulk}
\end{align}
where $K_{f_{AD}}, K_{f_{BD}}, K_{f_{CD}}$ are tensor products of Majorana operators $\gamma$ with appropriate phase factor. Again, this Hamiltonian leads to a trivial bulk wavefunction while the two-dimensional Majorana color code lives on boundaries of color $D$. The system needs to satisfy the fermionic parity symmetry as a whole, but each two-dimensional color code on the boundary may violate it, introducing an interesting anomaly. Indeed, the number of Majorana fermions on each boundary is odd for the choice of boundaries as in Fig~\ref{fig_color_boundary}. This three-dimensional Majorana color code may be useful for storing quantum information as the majorana edge mode on the boundary is encoded into the two-dimensional Majorana color code.

\subsection{Three-dimensional topological order}

Next, consider the following stabilizer Hamiltonian:
\begin{align}
H_{\text{bulk}} = - \sum_{f_{AB},f_{BC},f_{AC}} ( S^{X}_{f_{AB}} + S^{X}_{f_{BC}} + S^{X}_{f_{AC}} + S^{Z}_{f_{AB}} + S^{Z}_{f_{BC}} + S^{Z}_{f_{AC}}) \label{eq:bulk2}
\end{align}
where interaction terms live on $2$-cells of color $AB, AC, BC$. This Hamiltonian is equivalent to two copies of the three-dimensional toric code on a bulk. To see this, we should look for logical operators. There are $X$-type and $Z$-type string-like logical operators which connects $3$-cells of color D while there are $X$-type and $Z$-type membrane operators which anti-commute with string-like operators. 

Let us construct a Majorana version of the above Hamiltonian:
\begin{align}
\hat{H}_{\text{bulk}} = - \sum_{f_{AB},f_{BC},f_{AC}} ( K_{f_{AB}} + K_{f_{BC}} + K_{f_{AC}}).
\end{align}
The model supports a point-like fermionic excitation which can be created by a string-like Majorana operator. Since the statistics of excitations is identical to a stabilizer Hamiltonian model (the three-dimensional fermionic toric code) proposed by Levin and Wen~\cite{Levin:2003aa}, we expect that the above Majorana fermion model is identical to this model via suitable local relabelling of Majorana operators by Pauli operators. This procedure will be addressed elsewhere.

\bibliography{reference2018Jan, biblio_other}{}

\end{document}